\documentclass[fleqn,usenatbib,useAMS]{mnras}

\usepackage{lineno}
\usepackage{graphicx, caption}	
\usepackage{amsmath}	
\usepackage{amssymb}	
\usepackage{multicol}        
\usepackage{bm}		
\usepackage{pdflscape}	
\usepackage[table,xcdraw]{xcolor}
\usepackage{multirow}
\usepackage{longtable}
\usepackage{microtype}
\usepackage{url}

\usepackage[T1]{fontenc}
\usepackage{ae,aecompl}
\usepackage[utf8]{inputenc}
\usepackage{hyperref}
\DeclareUnicodeCharacter{2212}{-}
\DeclareUnicodeCharacter{03B3}{$\gamma$}
\DeclareUnicodeCharacter{03BD}{$\nu$}

\usepackage{newtxtext,newtxmath}

\title[Radio vs No  Radio]{Radio-bright vs. Radio-dark Gamma-ray Bursts - More Evidence for Distinct Progenitors. }

\author[]{Angana Chakraborty$^{1}$, Maria Dainotti$^{2,3,4}$, Olivia Cantrell$^{5}$, 
\newauthor Nicole Lloyd-Ronning$^{5,6}$\thanks{Corresponding author: lloyd-ronning@lanl.gov}
\\
$^{1}$Kolkata, India\\
\author[0000-0003-4442-8546]{M. G. Dainotti}
$^{2}$ National Astronomical Observatory of Japan, 2-21-Osawa, Mitaka, Tokyo, 181-8588, Japan \\
$^{3}$ Sokendai, Graduate University for Advanced Studies, Hayama, Miura district, Kanagawa 240-0193, Japan \\
$^{4}$ Space Science Institute, 4765 Walnut St., Boulder, CO, 80301, USA.\\
$^{5}$Los Alamos National Lab, Los Alamos, NM, USA 87545\\
$^{6}$University of New Mexico, 4000 University Dr., Los Alamos, NM, USA 87544
}

\pubyear{2022}

\begin{document}
\label{firstpage}
\pagerange{\pageref{firstpage}--\pageref{lastpage}}
\maketitle

\begin{abstract}
We analyze two distinct samples of GRBs, with and without radio afterglow emission. We use an updated sample of 211 GRBs and find, in agreement with previous results (although with a sample that is almost twice as large), that the intrinsic $\gamma$-ray duration ($T_{int}$) and isotropic equivalent energy ($E_{iso}$) distributions between these two populations appear to be significantly different. This implies that the radio-bright GRBs are more energetic and last longer than radio-dark GRBs. The two samples' redshift distributions ($z$) are not statistically different. We analyze several correlations between variables ($E_{iso}$, $T_{int}$, jet opening angle, and $z$), accounting for selection effects and redshift evolution using the Efron–Petrosian method. We find a statistically significant anti-correlation between the jet opening angle and redshift, as well as between $T_{int}$ and redshift, for both radio-bright and radio-dark GRBs. 
Finally, in agreement with previous work, we find that very high energy (0.1 − 100 GeV) extended emission is present only in the radio-bright GRB sample. Our work supports the possibility that the radio-bright and the radio-dark GRBs originate from different progenitors.

\end{abstract}

\begin{keywords}
stars(general)--gamma-ray bursts; cosmology
\end{keywords}




\section{Introduction} \label{introduction}

Gamma-ray bursts (GRBs) are the most energetic and luminous cosmological events, with prompt $\gamma$-ray emission lasting from ten milliseconds to several hours. The isotropic energy emitted in the $\gamma$-rays vary from $\sim 10^{48}$ to $\sim 10^{55}$ ergs \citep{CF2011}. Depending on the duration of their prompt emission, they are further classified into short (SGRBs) and Long GRBs (LGRBs) \citep{KM93}. SGRBs, whose prompt duration lasts less than two seconds, appear to originate from the merger of compact objects \cite{Ab17}. LGRBs, predominantly found in star-forming regions of late-type galaxies \citep{2004C, 2014L}, have prompt $\gamma$-ray emission lasting longer than $\sim 2 s$ \citep{GCG04} and likely originate from the death of massive, rapidly rotating stars \citep{P98, MW99, WH06}. 

For LGRBs and SGRBs, dissipation in the relativistic jetted outflow produces the prompt $\gamma$-ray emission through internal shocks or magnetic reconnection events. The subsequent long-lived afterglow emission (observed at X-ray, optical, and radio wavelengths) is produced through the interaction of the relativistic jetted outflow with the circumburst medium. For reviews summarizing these results, please see \cite{2004P, 2004ZM, 2006M, 2009G, 2015KZ, Lev16} and references therein.

In this paper, we focus on trends in LGRBs only. In particular, we examine the differences in properties and correlations present for those LGRBs with detected radio afterglows and compare them to those without detected radio afterglows.  The radio regime provides a unique window into probing the energetics and environment of the burst \citep{2000FW, 2008V}. 

\cite{2012CF} showed that for LGRBs, only about 31$\%$ have a detected radio afterglow (compared to the $\sim$ 95$\%$ and $\sim$ 70$\%$ of X-ray and optical afterglows, respectively) ($\sim$ 70$\%$), and attributed this lower detection rate to detector sensitivity.  However, \cite{2013HGM} suggested that some GRBs may be intrinsically radio-dark.
Examining this suggestion further, \cite{LR17, LR19} found that radio-bright GRBs are significantly longer in their prompt $\gamma$-ray duration and have higher isotropic energies. In addition, \cite{LR19} found a statistically significant anti-correlation (at the $> 4 \sigma$ level) between prompt duration and redshift in the radio-bright GRB population. \cite{2021Z} also reached similar conclusions. The authors suggested that radio-bright and radio-dark GRBs might arise from distinct progenitors. In particular,  \cite{LR22} suggested that radio-bright GRBs may originate from massive stars collapsing in interacting binary systems, while radio-dark GRBs may be from the collapse of isolated massive stars. They show that in certain cases (e.g., a massive star with a closely orbiting compact object companion), the binary interaction can allow for spin-up of the star, which provides a larger angular momentum reservoir for the resulting black hole-disk system after the collapse.  At the same time, the interaction (specifically tidal forces) leads to a more complex and dense environment around the binary system.  The former leads to a longer-lasting GRB jet (i.e., GRB prompt emission). The latter leads to an ISM-like medium and brighter radio emission (see Figures 2 and 3 of their paper for the predicted radio light curves in this model compared to a single-star model with a wind-like medium)\footnote{We note that the structure of the circumbinary medium, which can be very complex, manifests itself in practice as an approximately constant ISM-like medium in the GRB light curve; in other words, the density perturbations are too small scale to be visible in the GRB lightcurve \citep{NG07}.}.  Therefore, this model can account for the longer prompt duration and radio afterglow emission seen in the radio-bright subset of GRBs. Other considerations may lead to this radio-bright/radio-dark dichotomy depending on the source of the radio emission. 
 For example, if it primarily arises from the reverse shock (which depends on the magnetization of the jet, among other parameters), only progenitors and environments capable of creating strong, longer-lived reverse shocks will produce radio-bright GRBs. Reverse shock emission is useful in probing the GRB outflow towards the central engine, and modeling the observed emission will help constrain the magnetization of the ejecta. Alternatively, an extremely dense environment can cause synchrotron self-absorption of the radio light curve, leading to radio-dark GRBs. \cite{LR17} provides more discussion on these latter options. Other possibilities include a black-hole-driven engine being responsible for radio-bright GRBs while a magnetar-driven engine leads to detecting radio-dark GRBs \citep{2013HGM}. In this scenario, the underlying emission efficiency, related to the magnetic field strength (with less efficient emission for higher magnetic fields), produces the bi-modality; they argue that the magnetar field strength ultimately suppresses the radio emission.  Given the potential for this dichotomy to provide clues to underlying progenitors, it is worth exploring in more detail and carefully examining the differences in the observed properties of these sub-populations.
 
Regarding the physical interpretation of the radio-bright GRBs, \cite{2022LD} have investigated the applicability of the standard fireball model with a set of closure relationships (CRs) for a sample of 26 radio afterglows that display a clear break and a sub-sample of 14 GRBs that present a radio plateau. They test these samples against CRs corresponding to a constant-density interstellar medium (ISM) or a stellar wind medium in both fast- and slow-cooling regimes and several density profiles and consider sets of CRs both with and without energy injection. They discovered that 12 of the 26 GRBs (46\%), of which 7/12 present a radio plateau, fulfill at least one CR in the sets tested, suggesting that the data is largely incompatible with the standard fireball model. Of the fulfilled CRs, the most preferred environment is the ISM, SC, $\nu_{\rm m} < \nu < \nu_{\rm c}$ without energy injection. These results are consistent with previous studies \citep{KF20, Misra21} that test the standard fireball model via the CRs in radio. The latter paper is focused on the case of GRB 190114C. Although the current study does entail the difference between radio-dark and bright GRBs, it is nevertheless important to understand the properties of the radio-bright GRBs and their physical explanation to achieve more insights in the emission mechanism of radio afterglows.

Regarding the observational properties of GRBs, like any astrophysical observations, GRB flux measurements are subject to Malmquist bias. Some of their properties may also intrinsically evolve with cosmological redshift \citep{LR02, 2013DC, Dainotti2015, 2020LHA, Dainotti2022SNeIb}. These biases can affect the underlying correlation between observed variables of the GRB sample \citep{2011SN} and consequently lead to incomplete or incorrect conclusions about trends in the data. Hence, observations must account for this to get a handle on trends in the true underlying population. 

We use the extremely powerful non-parametric statistical technique, the Efron-Petrosian (EP) method, to obtain the intrinsic correlations between the parameters \citep{1992EP,1998EP}. Unlike the observed correlations, the corrected intrinsic correlations account for the selection bias due to instrumental thresholds and/or redshift and uncover true, underlying trends in the data. In our paper, we apply the EP method to relevant variables related to the GRB emission, such as the burst duration, $T_{int}$, the isotropic equivalent energy, $E_{iso}$, and the jet opening angle, $\theta_{j}$. Several works like \cite{1999LP, 2000LPref, LR02, 2013DC, 2015DDS, 2017DNM, 2017DNMb, LR19c,2020DLK, 2020bDMLS,  2020LHA, 2021DL, 2022LD, 2022DBL, 2022Quasars, 2022DNS} have recovered intrinsic relationships in many correlations with the use of this method. For a review of GRB correlations in the prompt and afterglow emission and selection biases related to those, refer to \cite{2017DDV,2018DDT, 2018DA}. 

We organize this paper as follows: in $\S$\ref{Data}, we describe our data sample, the formulation of the EP method, and its application to our sample. In $\S$\ref{results}, we present the results using the statistical techniques discussed in $\S$\ref{Data}. Finally, in $\S$\ref{discussion}, we briefly discuss the implication of the results in the context of different progenitors/environments and summarize our main conclusions.

\section{Data and Selection Effects} \label{Data}

\begin{figure}
    \centering
    \includegraphics[width=0.5\textwidth,height=4cm]{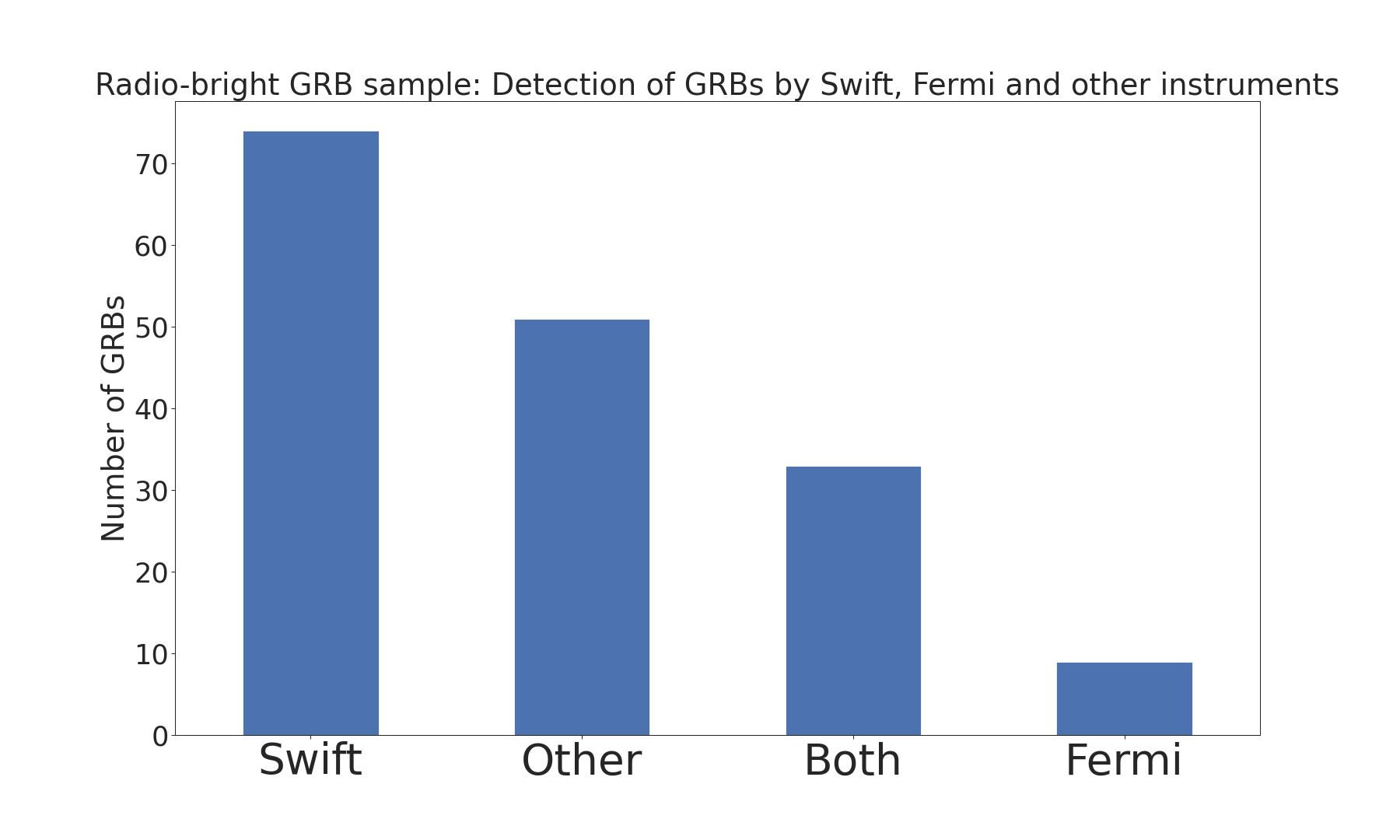}
    \includegraphics[width=0.5\textwidth,height=4cm]{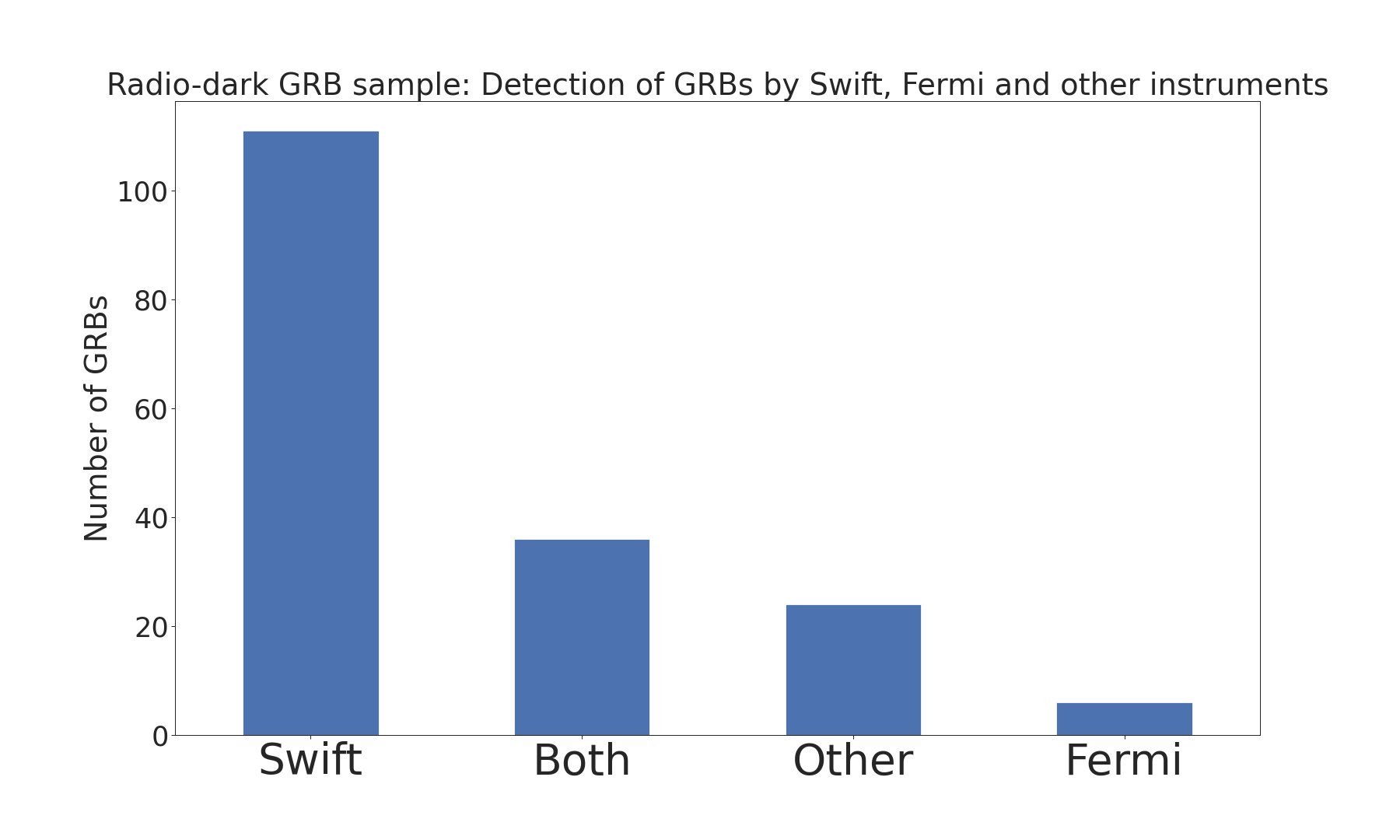}
    \caption{The number of radio-bright GRBs (upper panel) and radio-dark GRBs observed by Swift, Fermi, both of them and other instruments.}
    \label{fig:instruments}
\end{figure}

Our sample comprises 344 LGRBs, with prompt $\gamma$-ray flux primarily detected with the {\it Swift} satellite and with follow-up observations at radio wavelengths. Figure \ref{fig:instruments} shows the distribution of $\gamma$-ray detections by various satellites for our sample. We select our data sample from \cite{2012CF,  2020W,2021Z}, where 167 GRBs are detected in the radio and 177 are not. From this sample, we use only those GRBs which have measured redshifts. This leaves us with 211 GRBs - 123 GRBs with radio afterglow and 88 without.  
We note that the sample (and the radio detection rate) has considerably increased from \cite{2012CF}, while we have new facilities with more sensitivity; for example, we have an additional 139 detections from the Arcminute Microkelvin Imager (AMI) telescope. 
 We will refer to the GRBs with radio observations (radio flux density larger than 3$\sigma$ error bars) as radio-bright GRBs (see Table \ref{tab:radiobright}) and those without (with radio flux densities lower than 3$\sigma$ levels, including upper limits) as radio-dark GRBs from now onwards (see Table \ref{tab:radiodark}).



For our sample of GRBs with radio follow-up that has measured redshifts, we correct for cosmological time dilation and calculate the intrinsic duration of the burst as $T_{int} = T_{90}/(1 + z)$, where $T_{90}$ is the time interval over which a burst emits from 5$\%$ of its total measured counts to 95$\%$. Some bursts in the sample also have measured jet opening angles, $\theta_{j}$. We can use $\theta_{j}$ to calculate collimation-corrected energy $E_{\gamma}$, where $E_{\gamma}$ = $E_{iso} \cdot (1 - cos\theta_{jet}$). We note that  $E_{iso}$ is a reasonable estimate of the true emitted energy of the GRB prompt emission because it correlates strongly with the beaming-corrected energy \cite{Ghirland2007}. However, we also note that we find well-constrained jet opening angles for only a small fraction of GRBs in our sample. 

If we select GRBs whose $E_{iso} > 10^{52}$ erg, as done in \cite{LR19}, we will have a smaller total sample of 152 GRBs, where 94 GRBs are with a radio afterglow and 58 without a radio afterglow. This selection cut was an attempt by \cite{LR19} to mitigate sample bias by choosing only the brightest GRBs - those for which we might expect a radio afterglow according to the standard afterglow model. A crucial aspect of this study is to consider the potential biases in our sample and distinguish between those GRBs which are intrinsically radio-dark from those which are simply faint and fall below detector sensitivity limits. There are several possible ways to do this, including a more detailed examination of the broadband spectral and temporal differences between GRBs with and without radio afterglows and examining fluxes in other wavelength bands.  However, as mentioned in \cite{LR19}, there do not appear to be significant differences in the optical and X-ray spectral properties among radio-bright and dark samples to the extent the data are available.

\begin{figure*}
\begin{center}
\begin{multicols}{2}
   \hspace*{-3cm}
   \includegraphics*[width=0.65\textwidth,height=7cm]{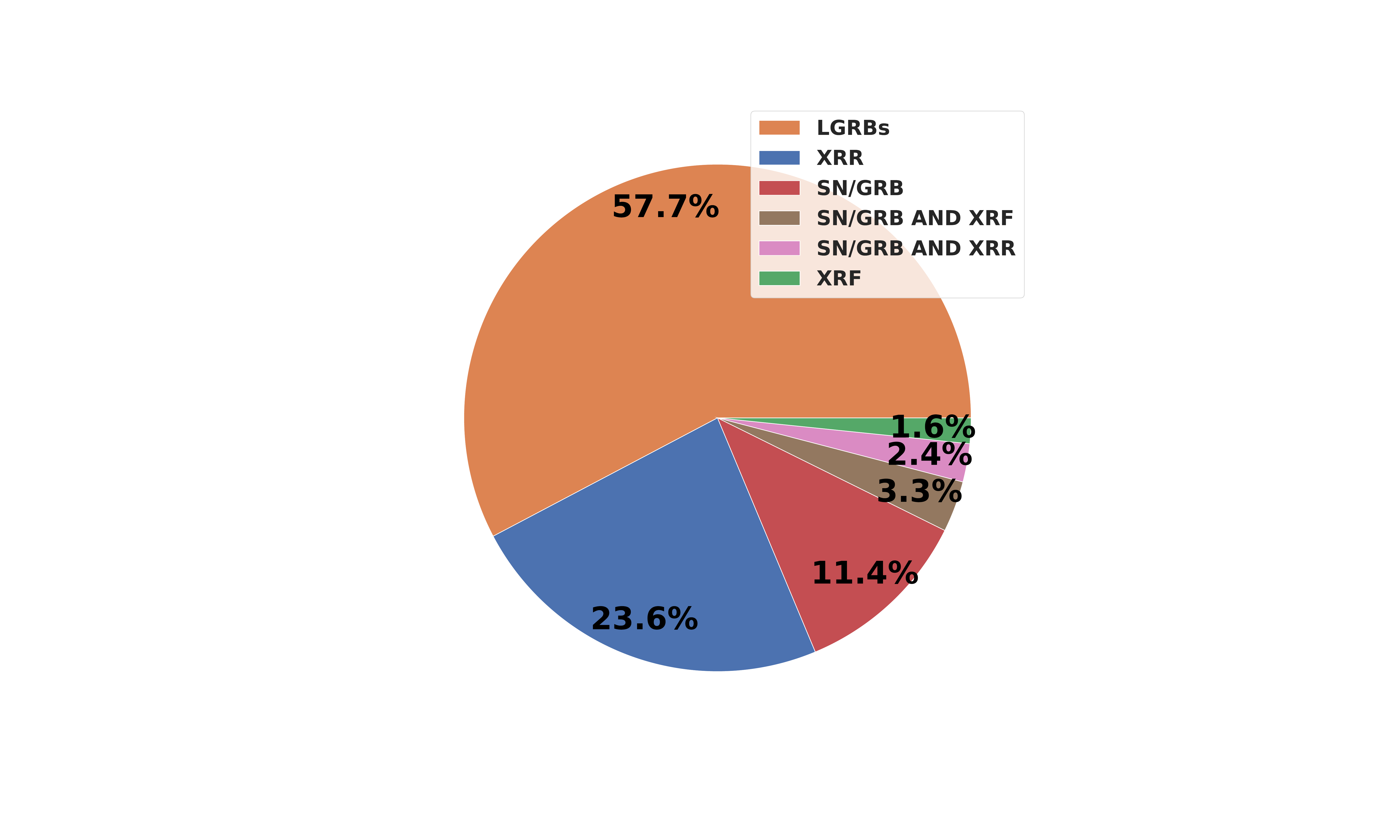}\par 
   \hspace*{-3cm}
   \includegraphics[width=0.65\textwidth,height=7cm]{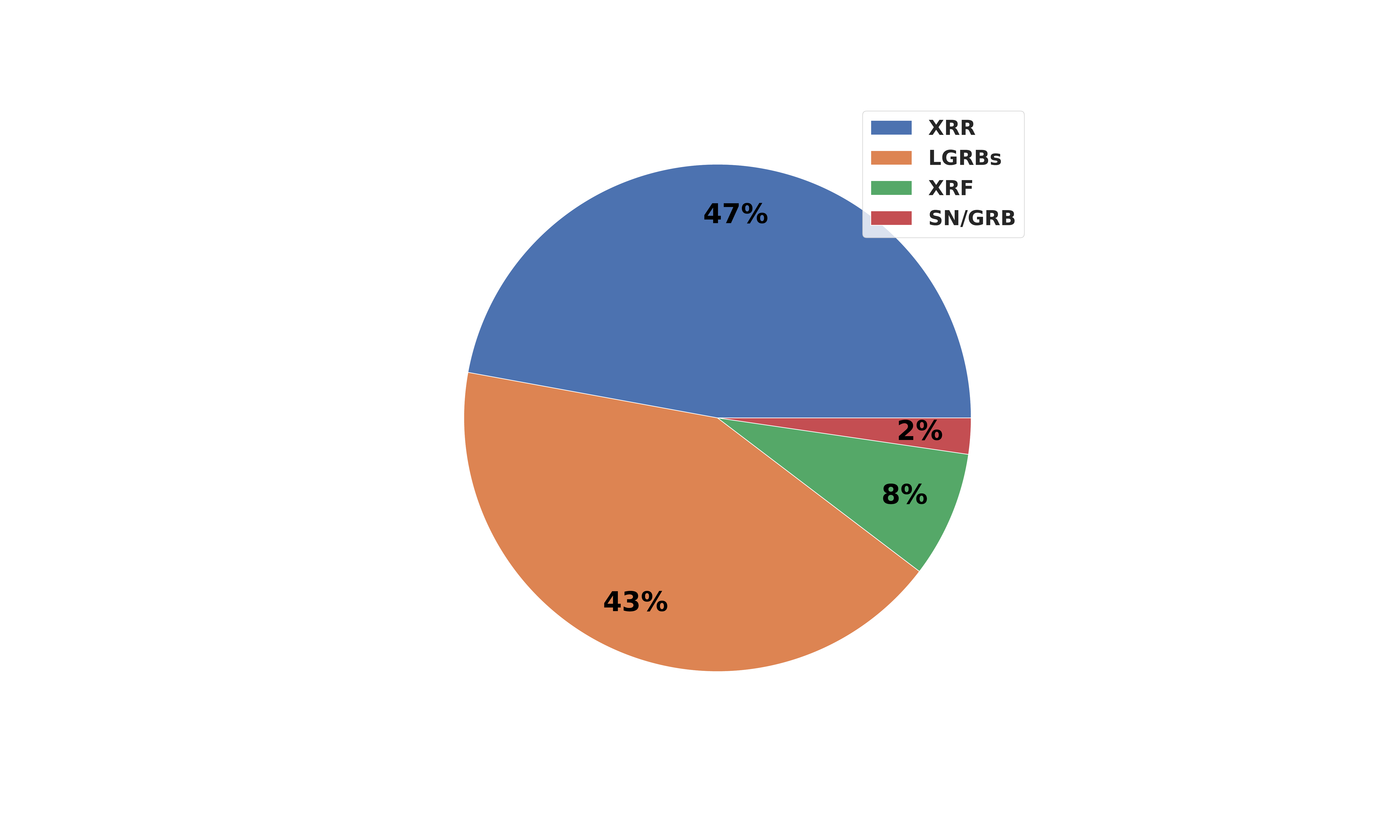}\par 
   \end{multicols}
   \vspace{-2cm}
\caption{The distribution of the different categories of LGRBS analyzed in our sample in a pie chart with their corresponding percentages. These categories include SN/GRB, XRF, and XRR. The left and right panels represent the radio-bright and radio-dark GRB samples, respectively.}
\label{fig:classes}
\end{center}
\end{figure*}

Although the isotropic energy cut ($E_{iso} > 10^{52}$) is a valuable one in examining radio-bright and dark GRBs, to study this more broadly dichotomy, we also examine a broader sample, which contains several previously defined subclasses of GRBs.
As pointed out in \cite{Dainotti2011, 2020DLAPJ, 2022LD} this is a key point in studying the GRB correlations and their physical interpretation.
To this end, our sample includes X-Ray Rich Gamma-Ray Burst (XRRs) and X-Ray Flashes (XRFs). X-ray flashes are classified as GRBs with X-ray fluence higher than the $\gamma$-ray fluence \citep{Heise2001}, while XRRs have stronger X-ray emission relative to most GRBs. However, it doesn't necessarily dominate the $\gamma$-ray emission \citep{2005B,2006DP,2008S,2018BM}. The general theory suggests that XRFs could be off-axis GRBs or maybe the tail end of the peak energy distribution of GRBs (for more details, please refer to \cite{2002YIN, 2003YIN, 2016SGP}). We also would like to point out that some GRBs show supernovae association, hereafter denoted with SN. Interestingly, over 90$\%$ of SN/GRB afterglows are radio-bright in our sample. 
We present the classification of our sample in Table \ref{tab:subclass} and Figure \ref{fig:classes}. 

\vspace{-5mm}

\subsection{Selection effects and correlation analysis} \label{technique}

To evaluate the intrinsic nature of correlations between relevant variables, we need to evaluate the intrinsic values of the slope of the correlations for our sample. In other words, it is crucial that we account for selection effects and artificial truncation due to detector sensitivity limits and non-independence of the variables (e.g., dependence on a particular variable with redshift). We use the Efron-Petrosian (EP) method to achieve unbiased correlations between the variables under consideration. The EP method is a powerful non-parametric statistical technique that can eliminate the selection effects of truncated data sets.

Here, we summarize the method briefly. The EP method uses a customized version of Kendall’s tau statistics to find out the best-fitting values of parameters describing the correlation functions. The test statistic, $\tau$, determines two variables' independence in a data set. The $\tau$ statistics definition reads as follows:
\begin{equation}
\tau = \frac{\sum_{i}({R_{i}  -  E_{i}})}{\sqrt{\sum_{i}{V_{i}}} }    
\end{equation}
where $E_{i} =  (1/2)(i + 1)$, $V_{i} = (1/12)(i^{2} − 1)$ are the expected mean, the variance and the rank $R_{i}$ in the associated sets which contain all GRBs with given defined threshold values.
A detailed description of the method is given in \cite{1992EP, 1999LP,2021DP} and references therein. In this paper, we use the EP method to determine, in particular, the impact of redshift evolution and selection bias on $E_{iso}$, $T_{int}$, and $\theta_{j}$.

We begin by first defining limiting values for all three variables. For $T_{int}$, the limiting values are given by $\frac{T_{min}}{1+z}$, where $T_{min}$ is the minimum observed time. For $E_{iso}$, we use the equation $E_{iso,lim} = 4\pi D^{2}_{L}(z)$; where $S_{lim}$ is the fluence limit and $D^{2}_{L}(z)$ is the luminosity distance assuming a flat $\Lambda$ CDM model with $\Omega_{M} = 0.3$ and $H_{0} = 70$ km $s^{−1}$  $Mpc^{−1}$. In all the cases, the limiting value should represent the population of data points and keep most of the sample size (at least 90$\%$, following the recipe of \cite{2013DC, 2015DDS}).

We initially followed the method given in \cite{2021DP} to find the limiting value. We compared the parent samples of all GRBs with known fluence values with smaller subsets of GRBs with known fluence and redshift values (refer Figure 1 in \cite{2021DP}). The fluences are between 15-150 keV and are in the units of $erg/cm^{2}$. We repeated the method for both radio-bright and radio-dark GRB samples. We made cuts in both the parent sample and the subsets at regular intervals and designated limiting fluences at those positions. For every cut, we considered the data above the limiting fluence value. We estimated the geometric distance between the two samples and the probability that the subset of GRBs was drawn from the parent sample by using the two-sample KS test between the data of the parent sample and the data sample above the cut. We selected the limiting fluence such that the probability that the parent population and the chosen sample are from the same distribution is approximately 1.0. Again, this method was used to check if the limiting values we used for all three variables, $T_{int}$, $E_{iso}$ and $\theta_{j}$, have a probability of $\sim 1$, implying that the subsamples of both radio-dark and bright burst (above the limiting line) were drawn from the same parent sample. 
Having established that the limits are well posed, we also checked that these choices retain at least 90$\%$ of the total sample for each variable. 

\section{Results} \label{results}

\subsection{Distribution of GRB parameters}
\begin{figure}
    \vspace{-20mm}
    \includegraphics[width=0.5\textwidth,height=20cm]{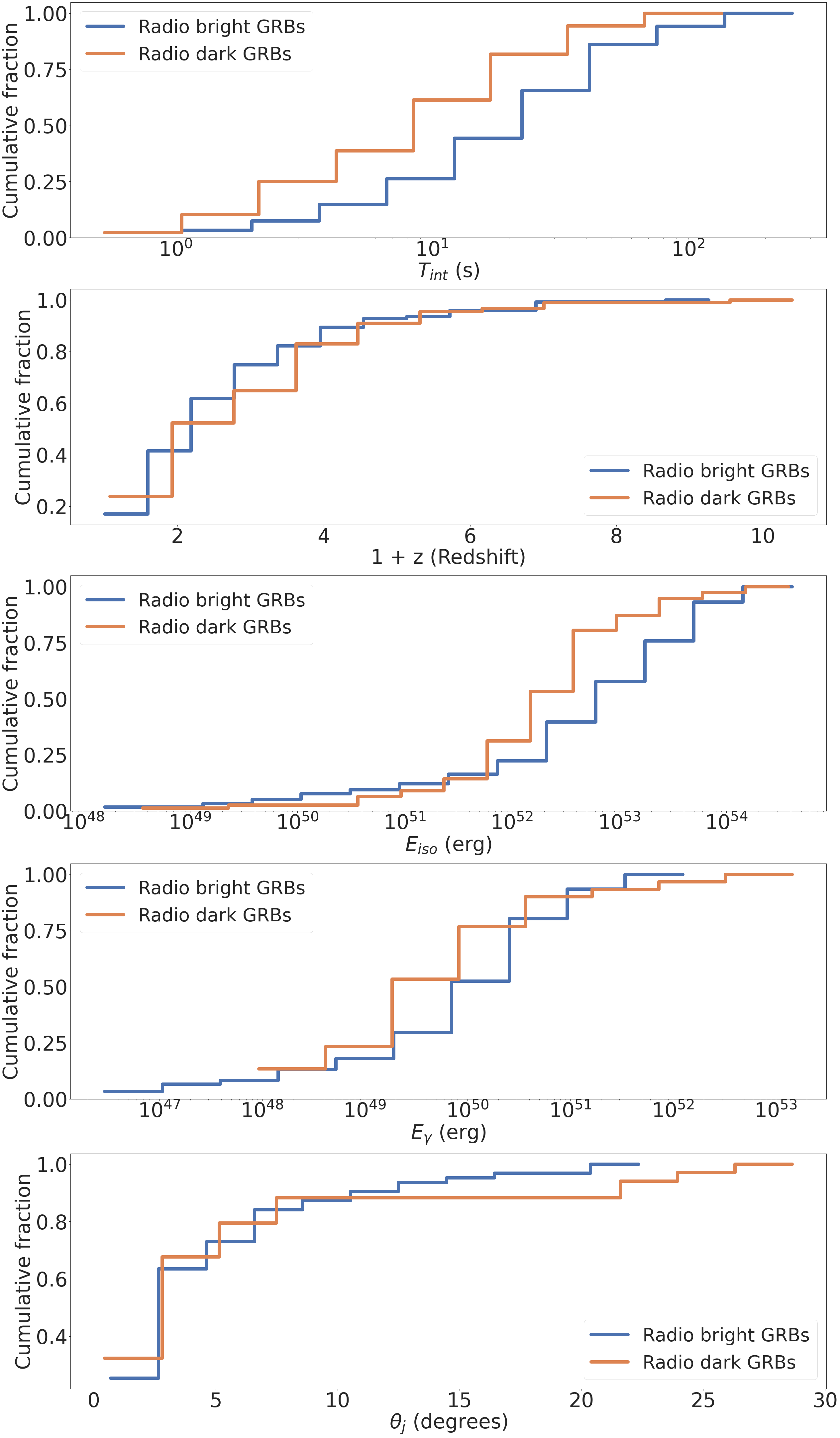}
    \vspace{-5mm}
    \caption{The cumulative distributions of (from top to bottom) the intrinsic duration $T_{int}$, redshift (1 + $z$), the isotropic equivalent energy $E_{iso}$, the collimation corrected energy, $E_{\gamma}$, and the jet opening angle $\theta_{j}$ for the radio-bright (in blue) and radio-dark (in red) GRBs.}
    \label{fig:CD_all}
\end{figure}

\begin{figure}  
    
    \includegraphics[width=.49\textwidth,height=4.6cm]{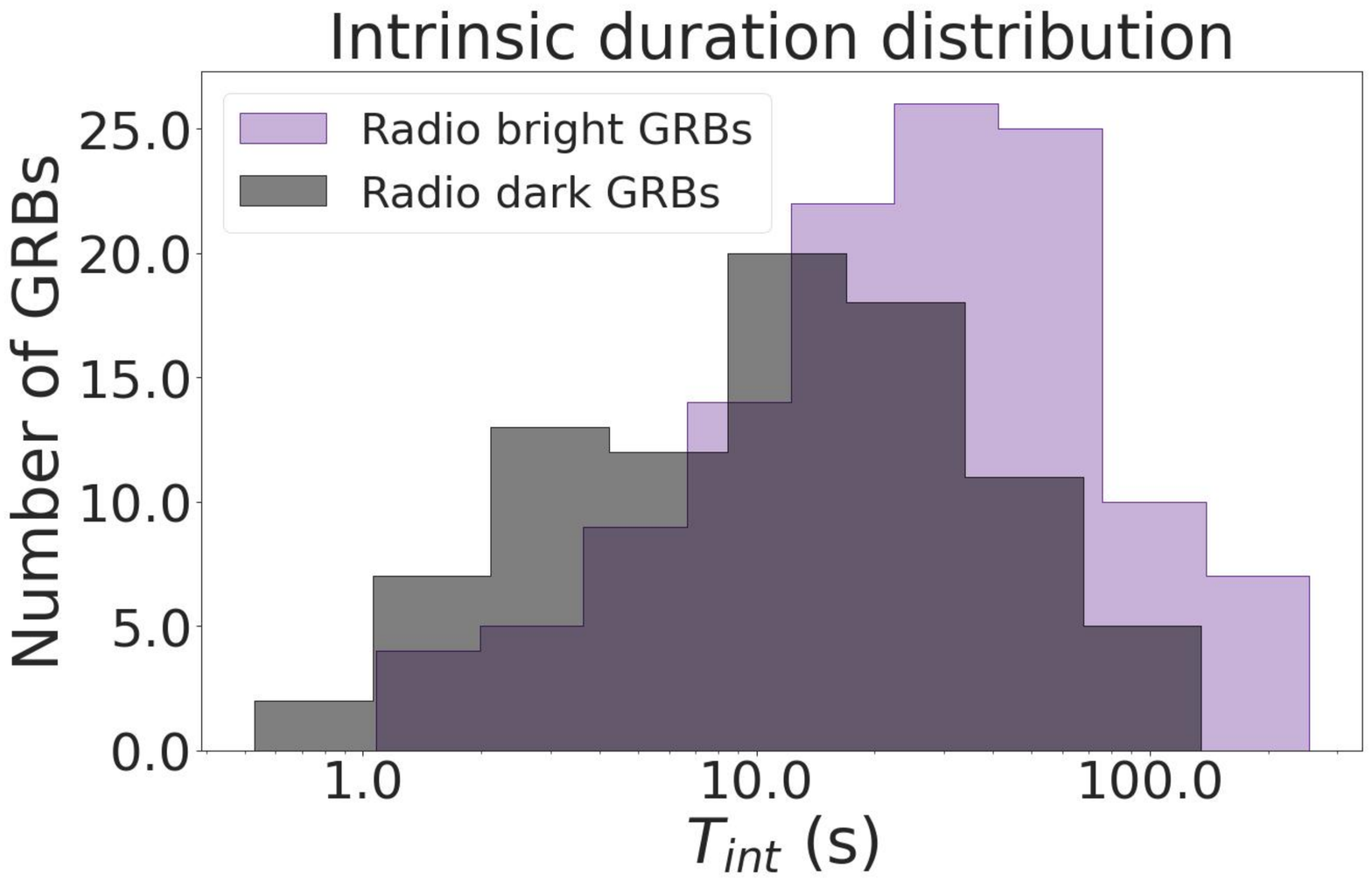}
    
    \includegraphics[width=.49\textwidth,height=4.6cm]{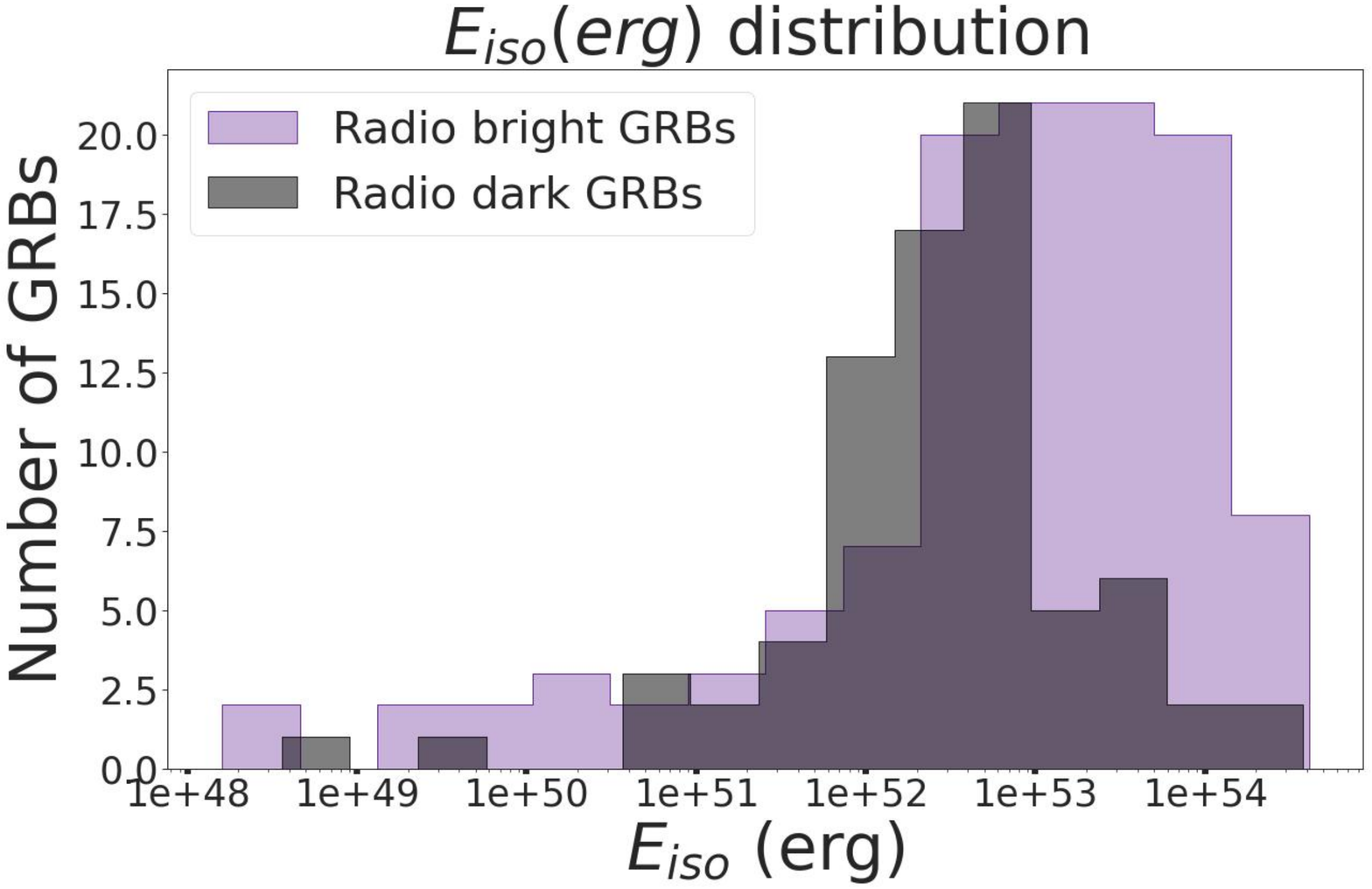}
    
    \includegraphics[width=.49\textwidth,height=4.6cm]{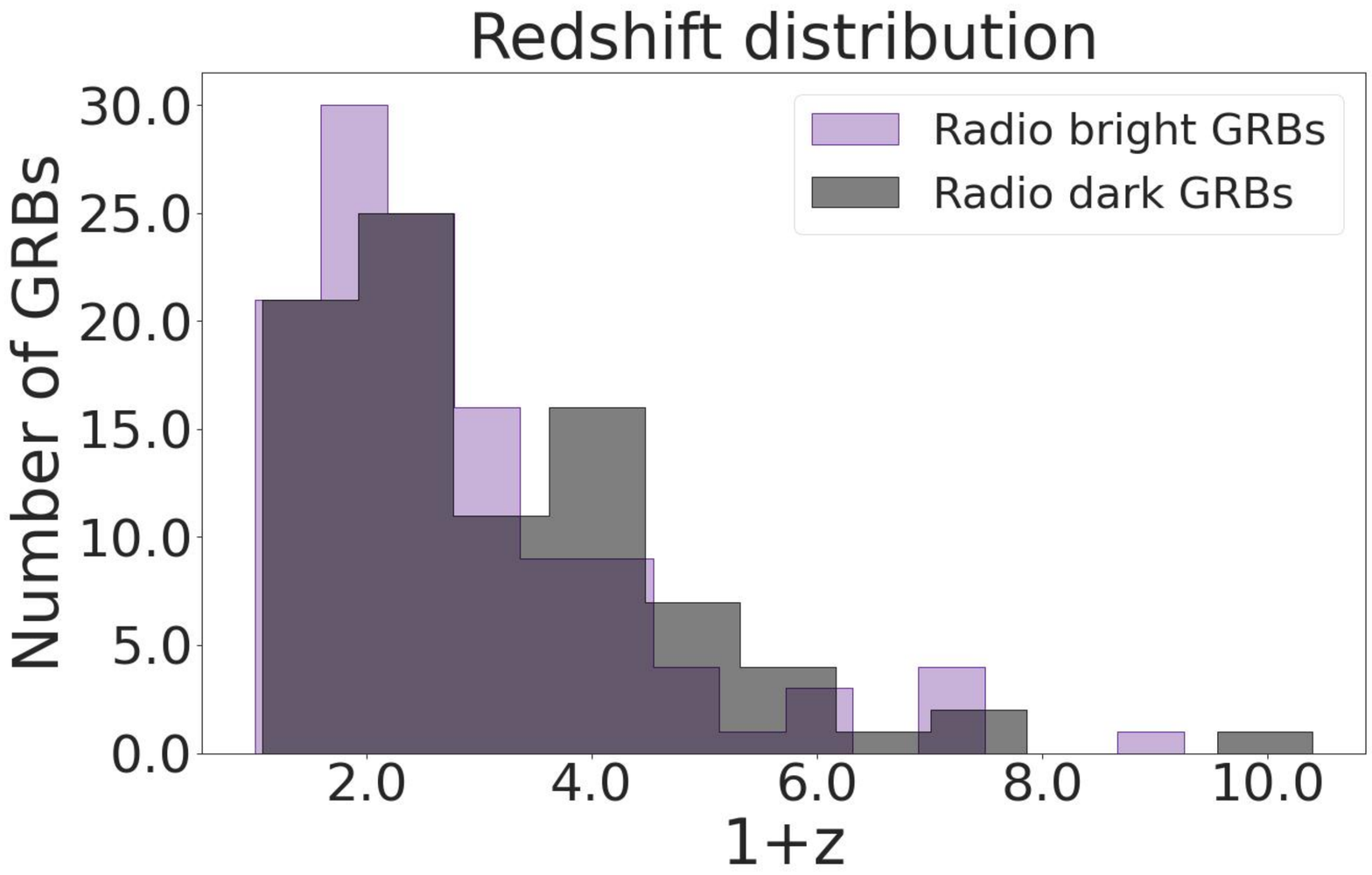}
    
    \includegraphics[width=.49\textwidth,height=4.6cm]{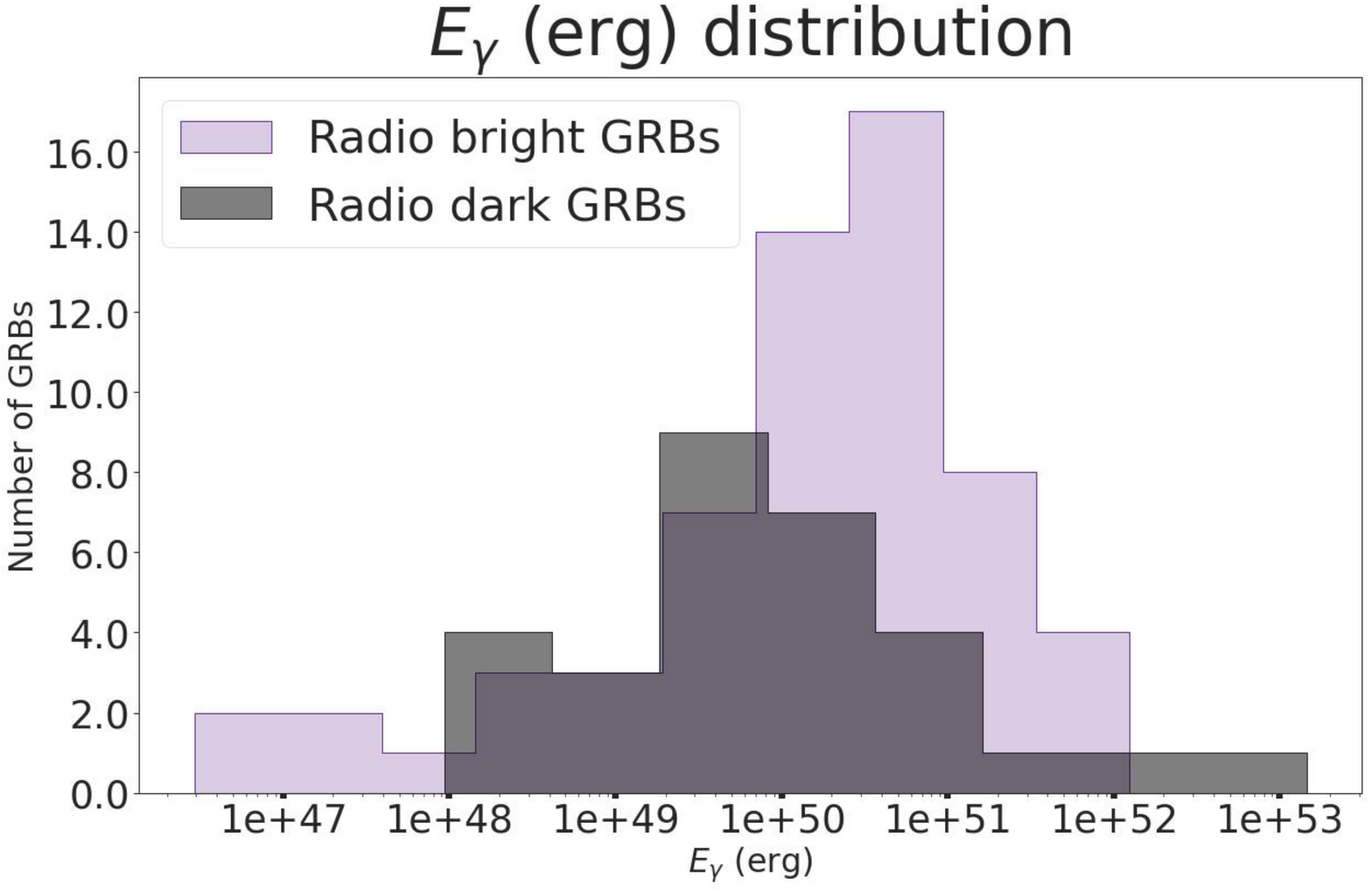}
    
    \includegraphics[width=.49\textwidth,height=4.6cm]{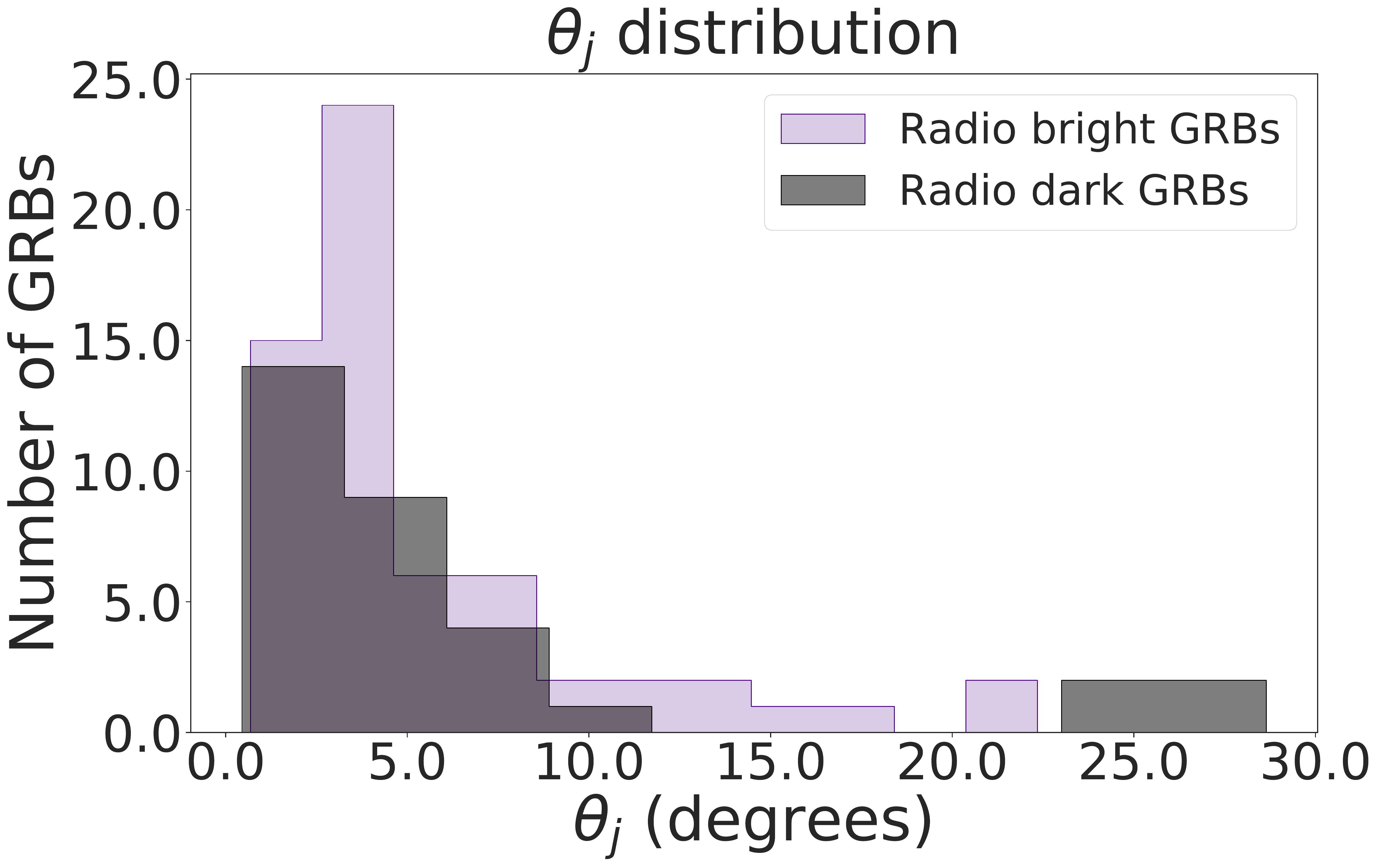}
    \vspace{-6mm}
    \caption{Differential distributions of the $T_{int}$, $E_{iso}$, redshifts, $E_{\gamma}$ and $\theta_{j}$ of radio-bright (purple) and radio-dark (grey) GRBs of our sample.}
    
    \label{histograms}
\end{figure}

For our sample of 211 GRBs, we plot the distributions of $z$, $E_{iso}$, $T_{int}$, $E_{\gamma}$ and $\theta_{j}$, for radio-bright and radio-dark GRBs.  Figure \ref{fig:CD_all} shows the cumulative distributions, while Figure \ref{histograms} shows the differential distributions of the various properties. We give the mean values of different parameters in Table \ref{tab:meanvalues}.  We investigate the radio-bright and radio-dark samples in the following sections to explore statistically whether they potentially have different origins based on their observational properties.

\subsection*{Redshift, isotropic equivalent energy, and intrinsic duration:} 
\begin{figure}
    \includegraphics[width=0.5\textwidth,height=6cm]{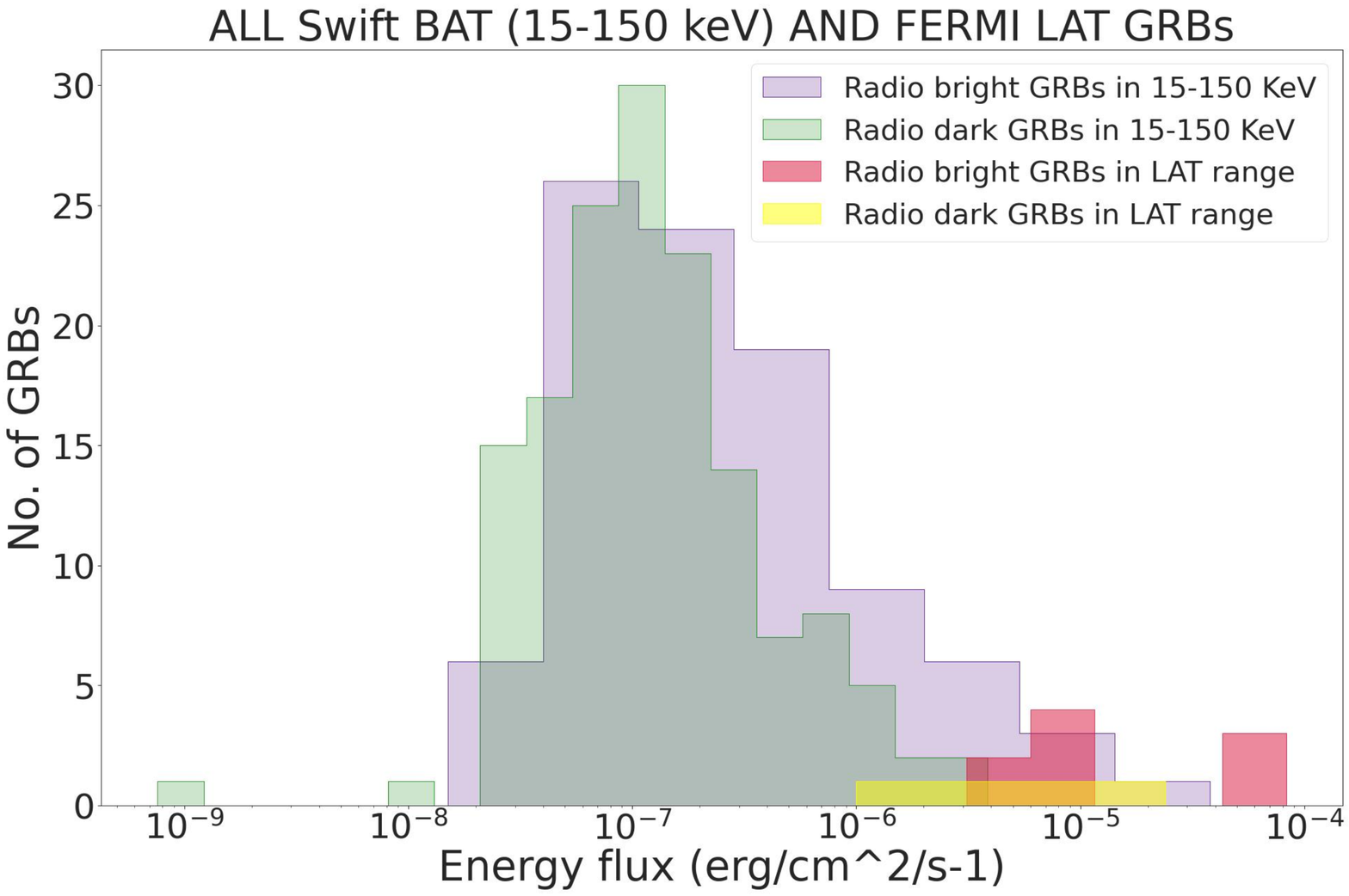}
    \vspace{-6mm}
    \caption{The differential distribution of the energy fluxes ($erg \: cm^{-2} s^{-1}$) in the $\gamma$-rays band and in the Swift band.}
    \label{fig:gammabandfluxes}
\end{figure}

Figure \ref{histograms} and Table \ref{tab:stat_tests} confirm that the $T_{int}$ of radio-bright GRBs are significantly longer than the radio-dark GRBs, with the mean value of $T_{int}$ of radio-bright GRBs is twice as long as the mean value of radio-dark GRBs (see Table \ref{tab:meanvalues}). We also see higher $E_{iso}$ in the radio-bright sample compared to the radio-dark sample in agreement with \cite{LR17, LR19, 2021Z}. Radio-bright GRBs also show higher fluxes in the $\gamma$-ray band than radio-dark GRBs (Figure \ref{fig:gammabandfluxes}).

The Student’s t-test yields a probability of more than 4 $\sigma$ confidence that the difference in the mean of $T_{int}$ between the radio-dark and radio-bright samples means is statistically significant (Table \ref{tab:stat_tests}). 

Regarding $E_{iso}$, a Student’s t-test yields only a moderate probability that the mean of the radio-dark and radio-bright $E_{iso}$ distributions are different and do not yield a significant difference between the two sample means for the redshift distribution, $z$.

Meanwhile, a KS test between the two samples yields a significant difference ($p < 10^{−3}$, where $p$ or the p-value is the probability that we cannot reject the hypothesis that the two distributions are the same) for both the $T_{int}$ and $E_{iso}$ distributions. The $z$ distribution is not statistically different (Table \ref{tab:stat_tests}). This supports the claim made in \cite{LR19} that the $T_{int}$ and $E_{iso}$ of the radio loud and quiet samples are significantly different, while the $z$ distribution is similar for both samples (although we point out again that their sample comprised only those GRBs with isotropic equivalent energies above $10^{52}$ erg). 

\begin{figure}
    \vspace{-20mm}
    \includegraphics[width=0.5\textwidth,height=20cm]{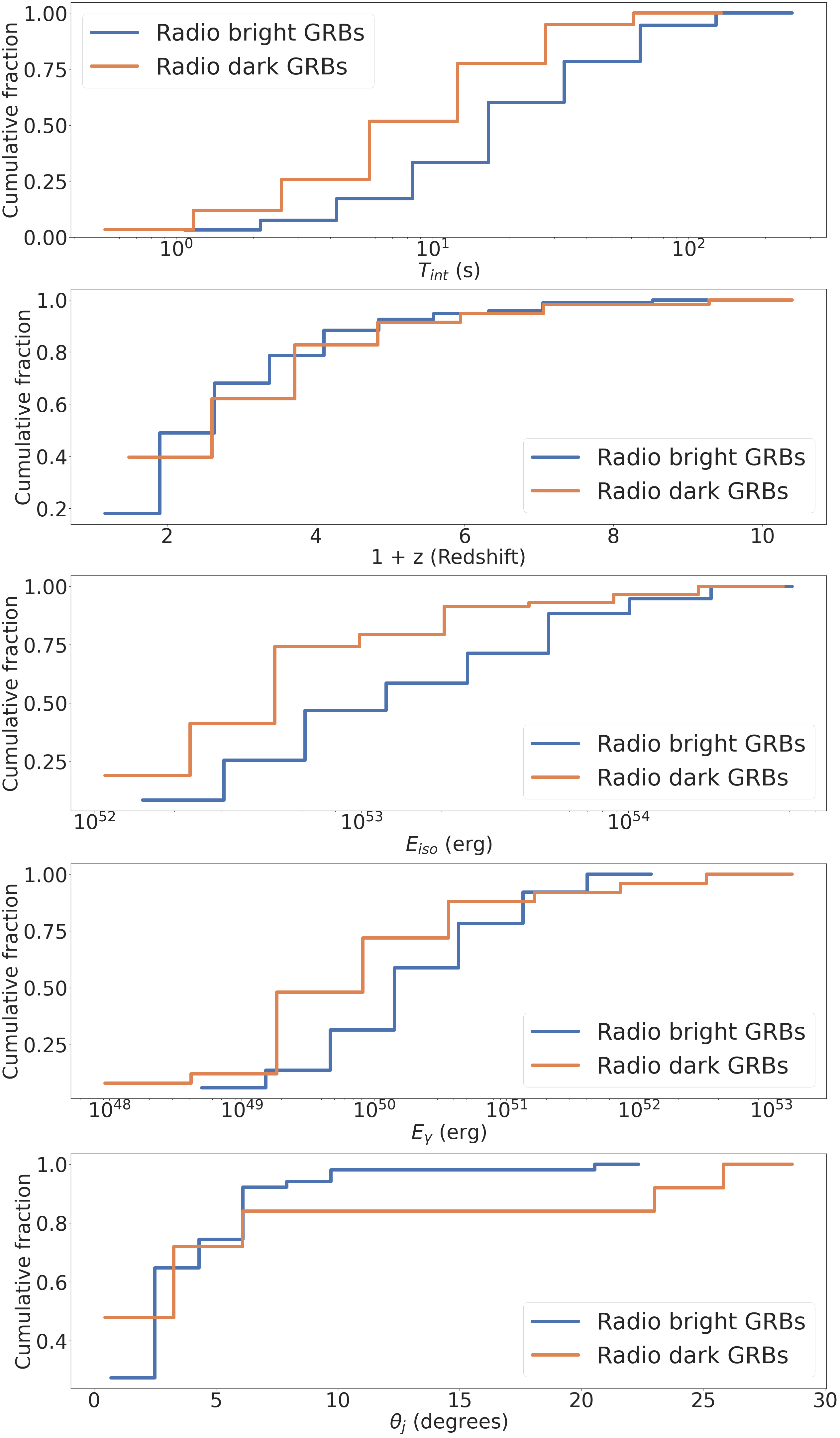}
    \vspace{-5mm}
    \caption{Cumulative distributions of intrinsic duration $T_{int}$, (1 + $z$), the isotropic equivalent energy $E_{iso}$, and the collimation corrected energy $E_{gamma}$ for the radio-bright (in blue) and radio-dark (in red) GRBs with $E_{iso}$ > $10^{52} erg$.}
    \label{fig:CD_cut}
\end{figure}

\begin{figure}      
    \includegraphics[width=.49\textwidth,height=4.6cm]{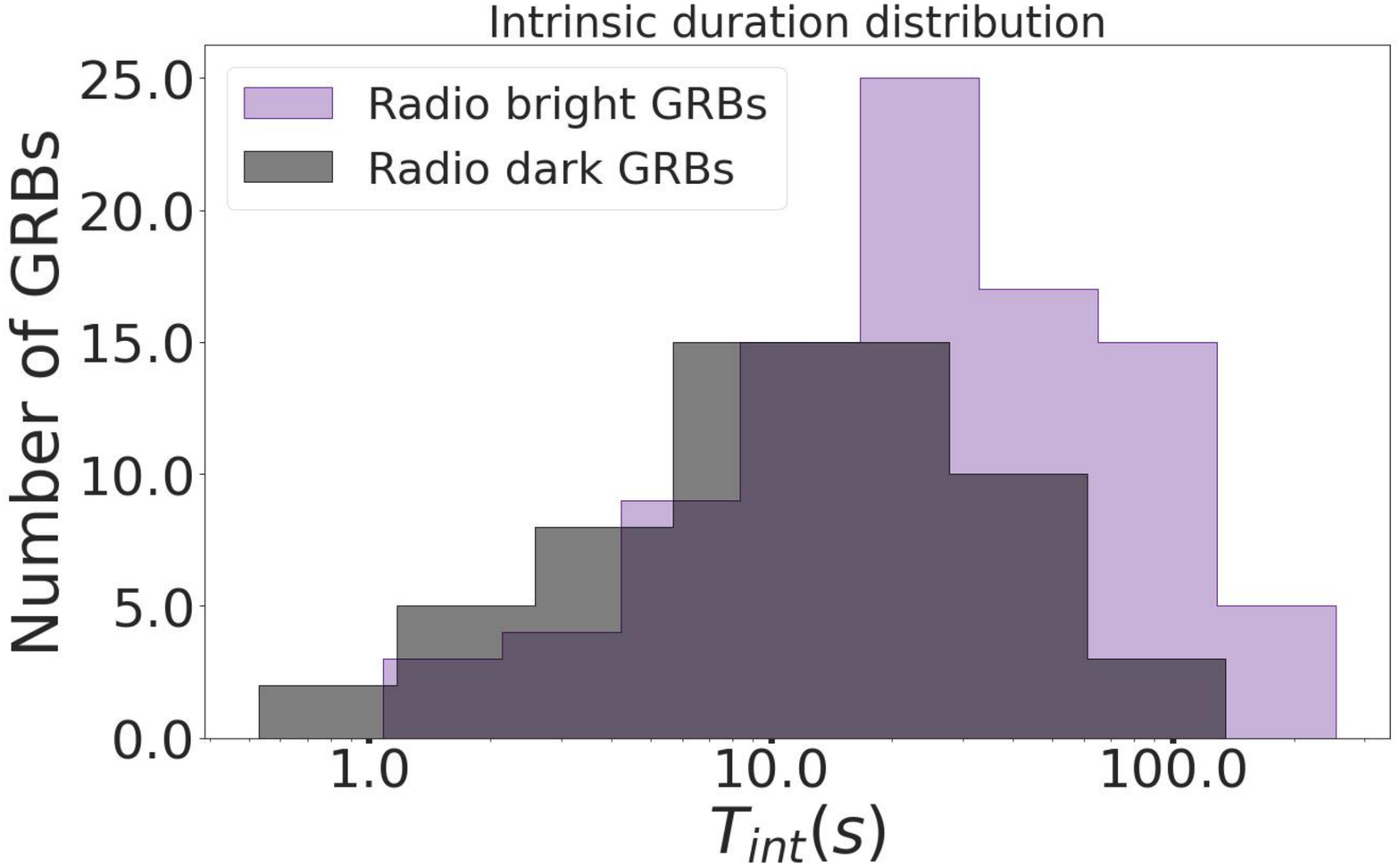}
    \vspace{1mm}
    \includegraphics[width=.49\textwidth,height=4.6cm]{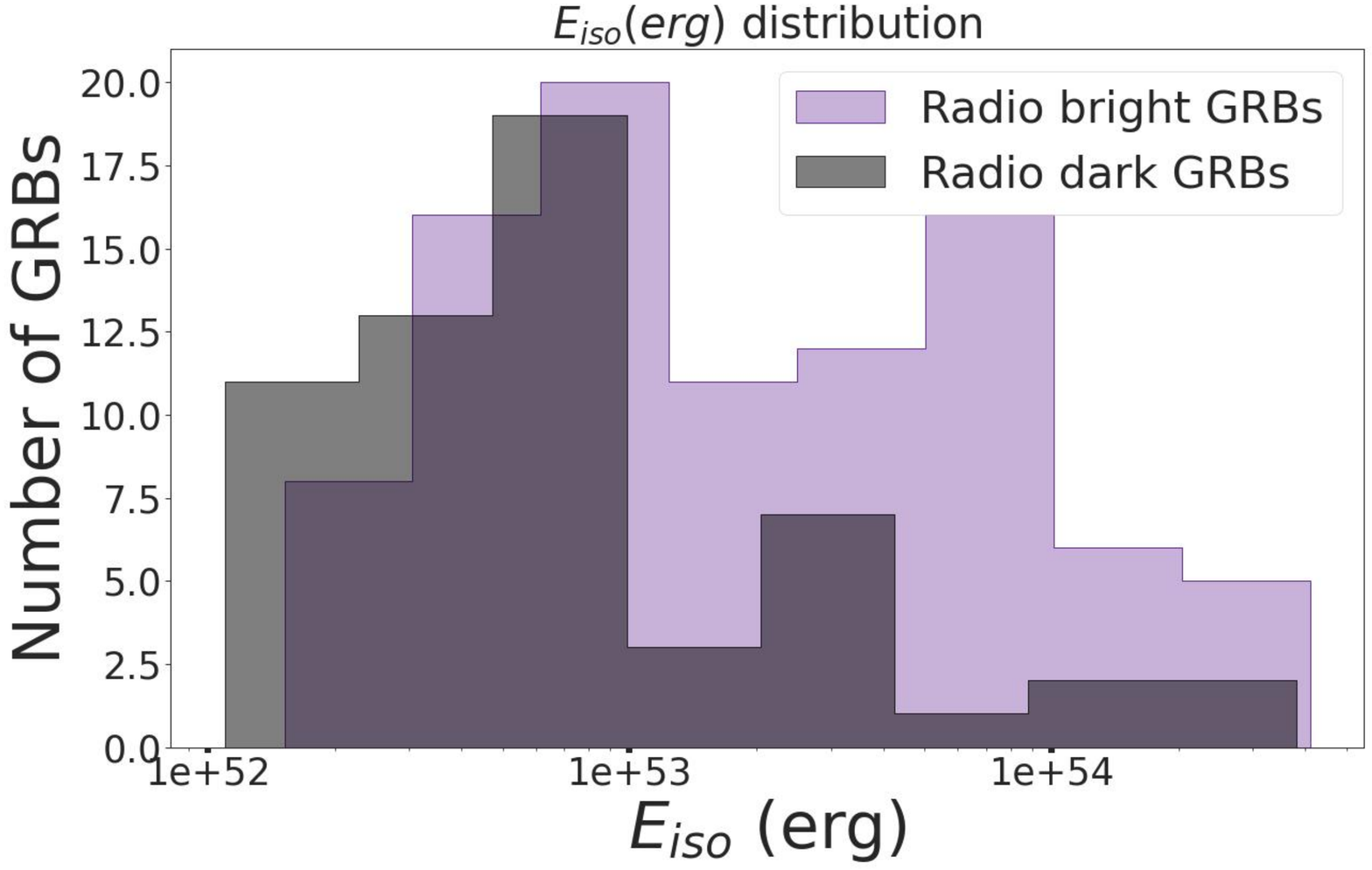}
    \vspace{1mm}
    \includegraphics[width=.49\textwidth,height=4.6cm]{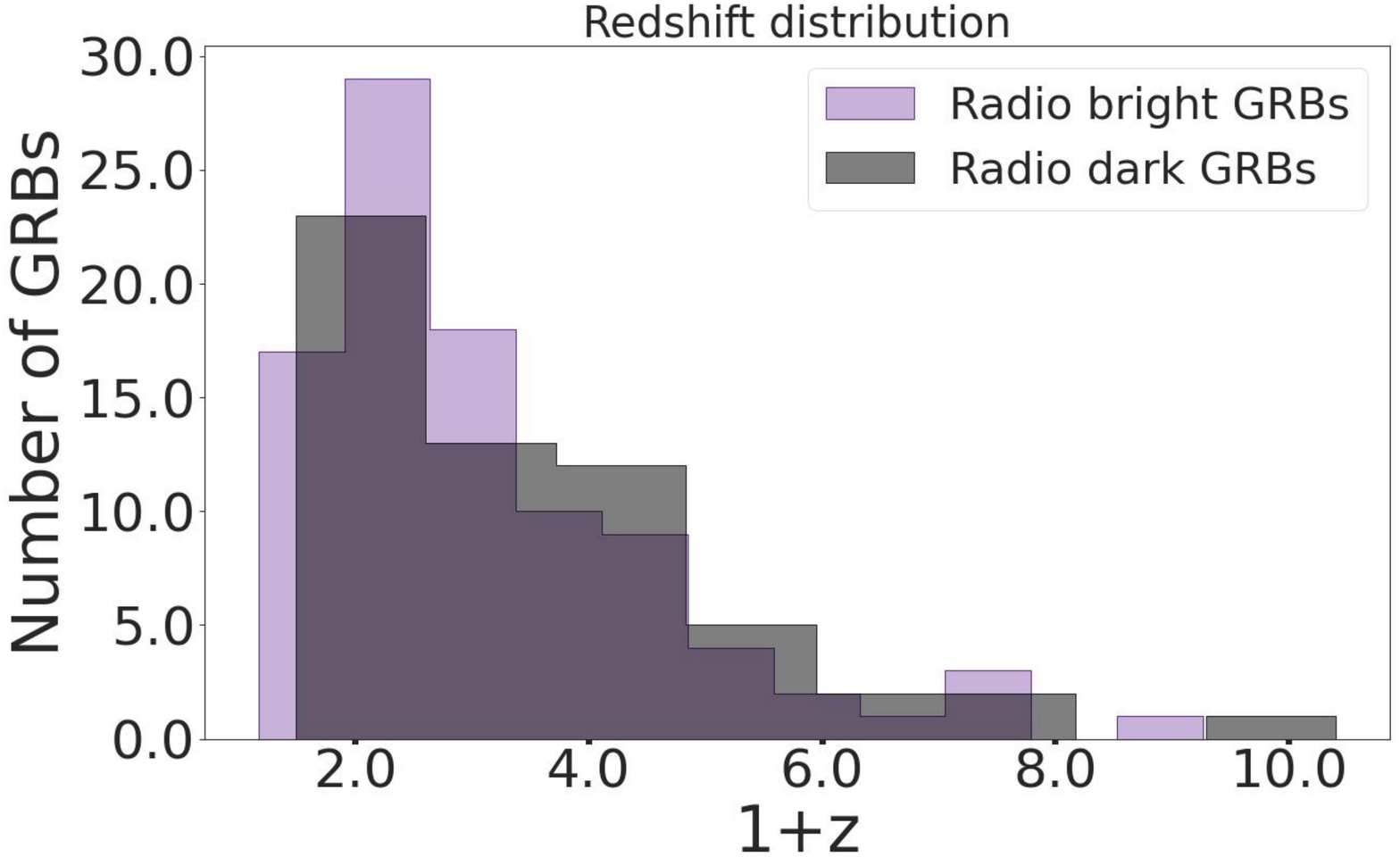}
    \includegraphics[width=.49\textwidth,height=4.6cm]{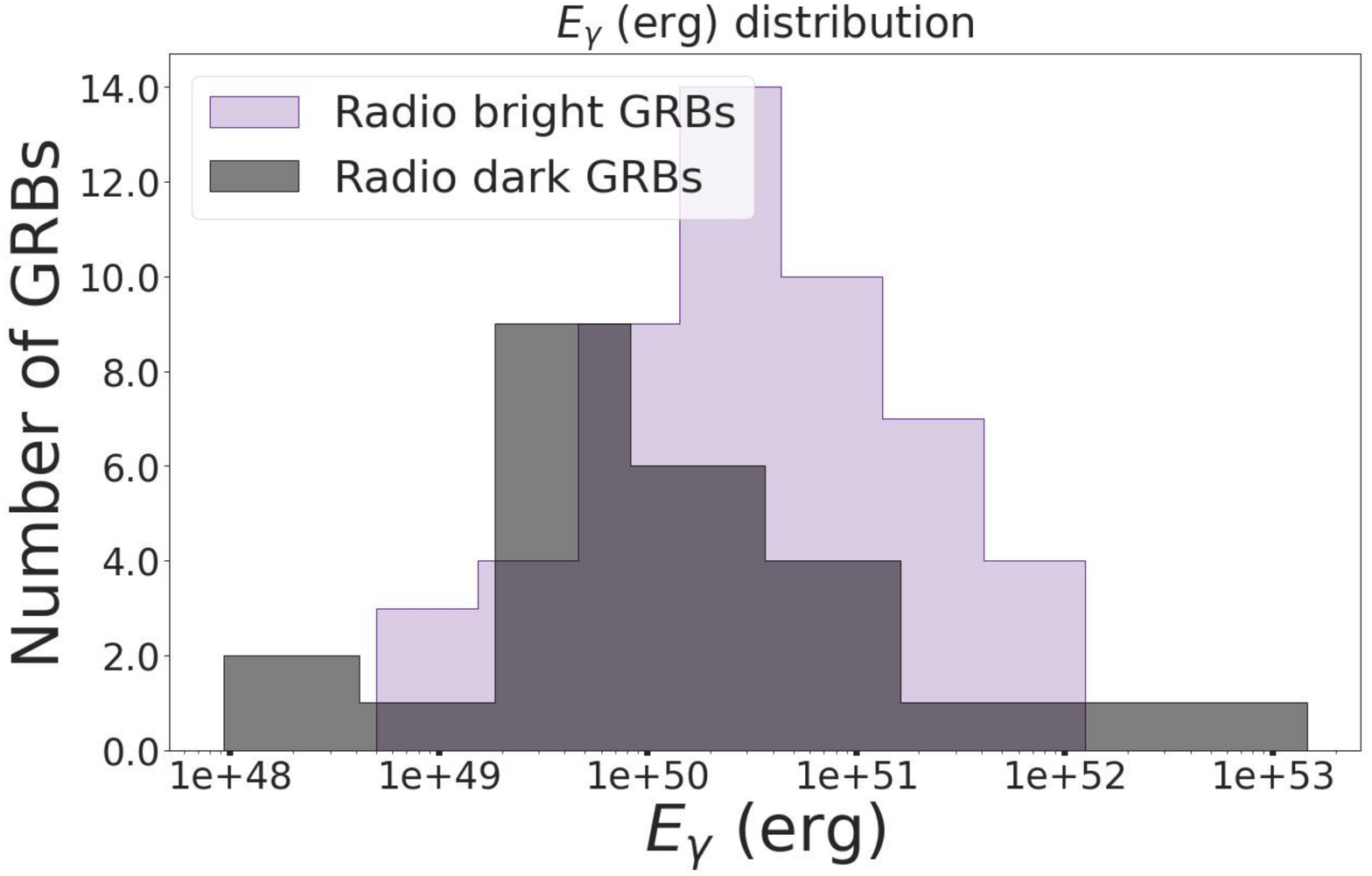}
    \vspace{1mm}
    \includegraphics[width=.49\textwidth,height=4.6cm]{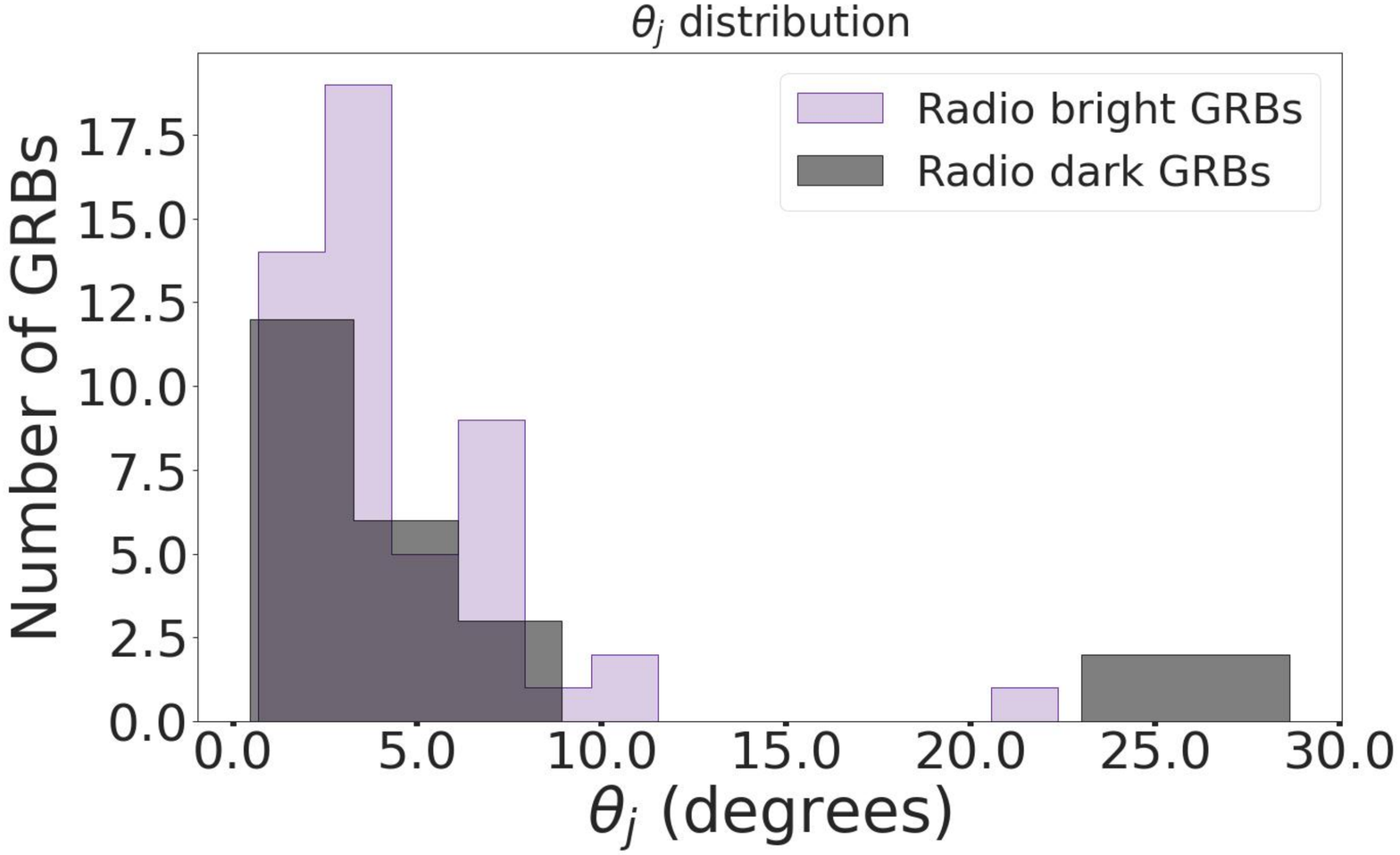}
    \vspace{-3mm}
    \caption{Differential distributions of the $T_{int}$, $E_{iso}$, redshifts, $E_{\gamma}$ and $\theta_{j}$ of radio-bright and radio-dark GRBs with $E_{iso} > 10^{52} erg$.}
    \label{histograms_cutEiso}
\end{figure}

If we constrain $E_{iso}$ and select only those GRBs with $E_{iso}$ > $10^{52} erg$, as done in \cite{LR19}, we obtain a subset of 94 GRBs with radio afterglow and 58 GRBs without radio afterglow. We show the comparison of the cumulative distributions of different GRB parameters for radio-bright and radio-dark GRB samples in Figure \ref{fig:CD_cut}, and show the results of the Student's t-test and KS tests for the redshift, $T_{int}$ and $E_{iso}$ distributions of this subset in Table \ref{tab:stat_tests_Eisocut} (see Figure \ref{histograms_cutEiso} for differential distributions of the $T_{int}$, $E_{iso}$, redshifts, $E_{\gamma}$ and $\theta_{j}$).  We find similar results to what was found above without this cut - the mean of the intrinsic duration distribution is significantly different, with radio-bright GRBs being, on average significantly longer in their prompt duration than radio-dark GRBs.  A KS test yields the distributions of $T_{int}$ and $E_{iso}$ distributions are significantly different, but the redshift distributions are not.

\subsection*{Jet opening angles and collimation corrected energies}
Only a fraction of GRBs has known jet opening angle values, as measured by achromatic steepening in the light curve. This is not surprising since it is difficult to constrain $\theta_{j}$, and jet break measurements are fraught with uncertainties. 
Table \ref{tab:stat_tests_jetangles} shows the results of Student's t-tests and KS tests for this sub-sample with jet opening angles (again, we only consider those GRBs with measured redshifts).  There is not a particularly convincing dichotomy between the radio-bright and radio-dark GRBs in this sub-sample, although the small numbers make it difficult to assess this statement.  The jet opening angle distributions, in particular, do not appear to be significantly different between radio-bright and radio-dark GRBs.

\begin{figure}
\includegraphics[width=.5\textwidth,height=6cm]{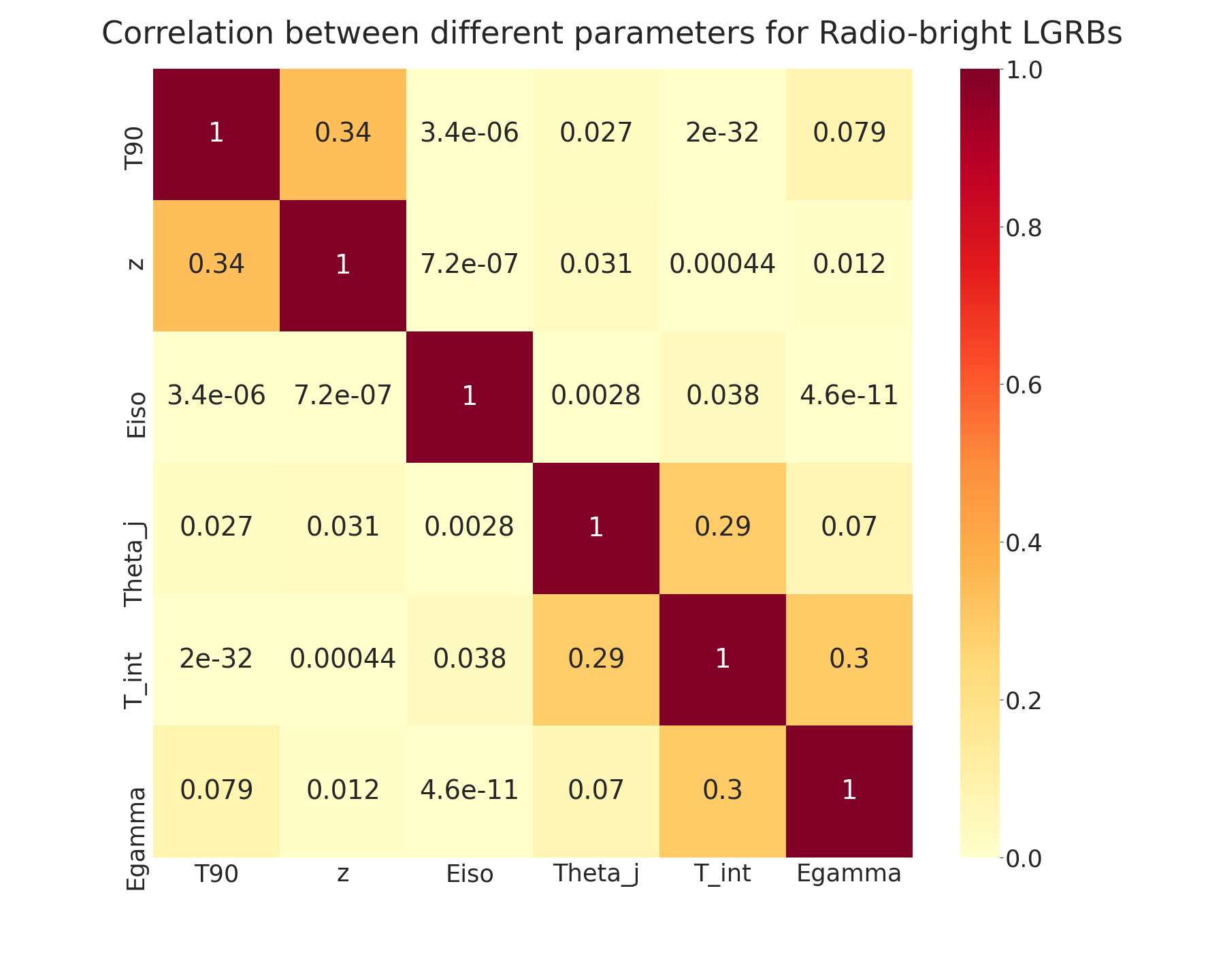}\hfill
\includegraphics[width=.49\textwidth,height=6cm]{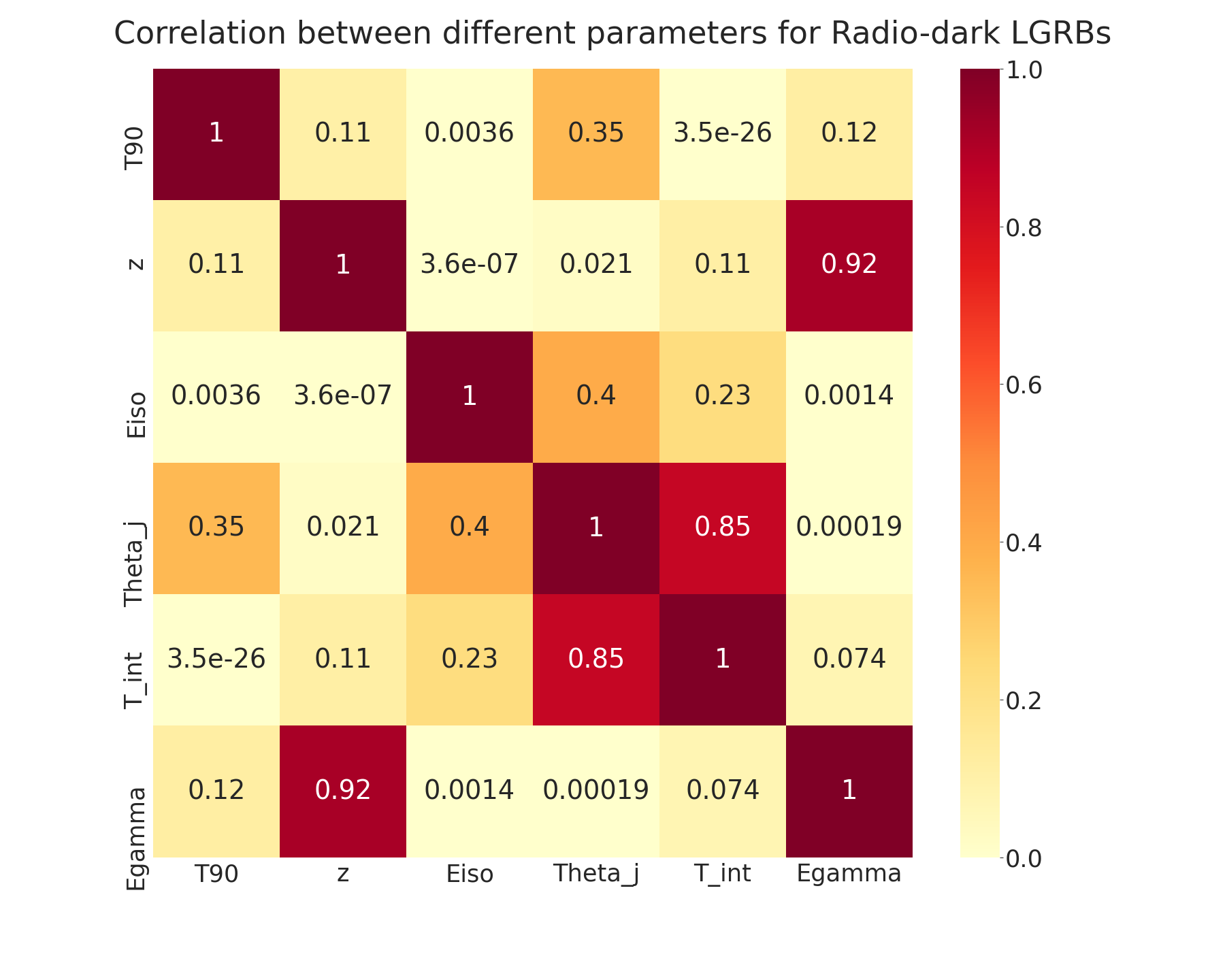}
\caption{The Kendall rank correlation between different parameters for radio-bright (upper panel) and radio-dark (lower panel) GRB sub-populations before applying the EP method.}
\label{correlation}
\end{figure}

\subsection{Correlation results}
\begin{figure}
 \hspace{-0.5cm}   \includegraphics[width=\linewidth,height=6cm]{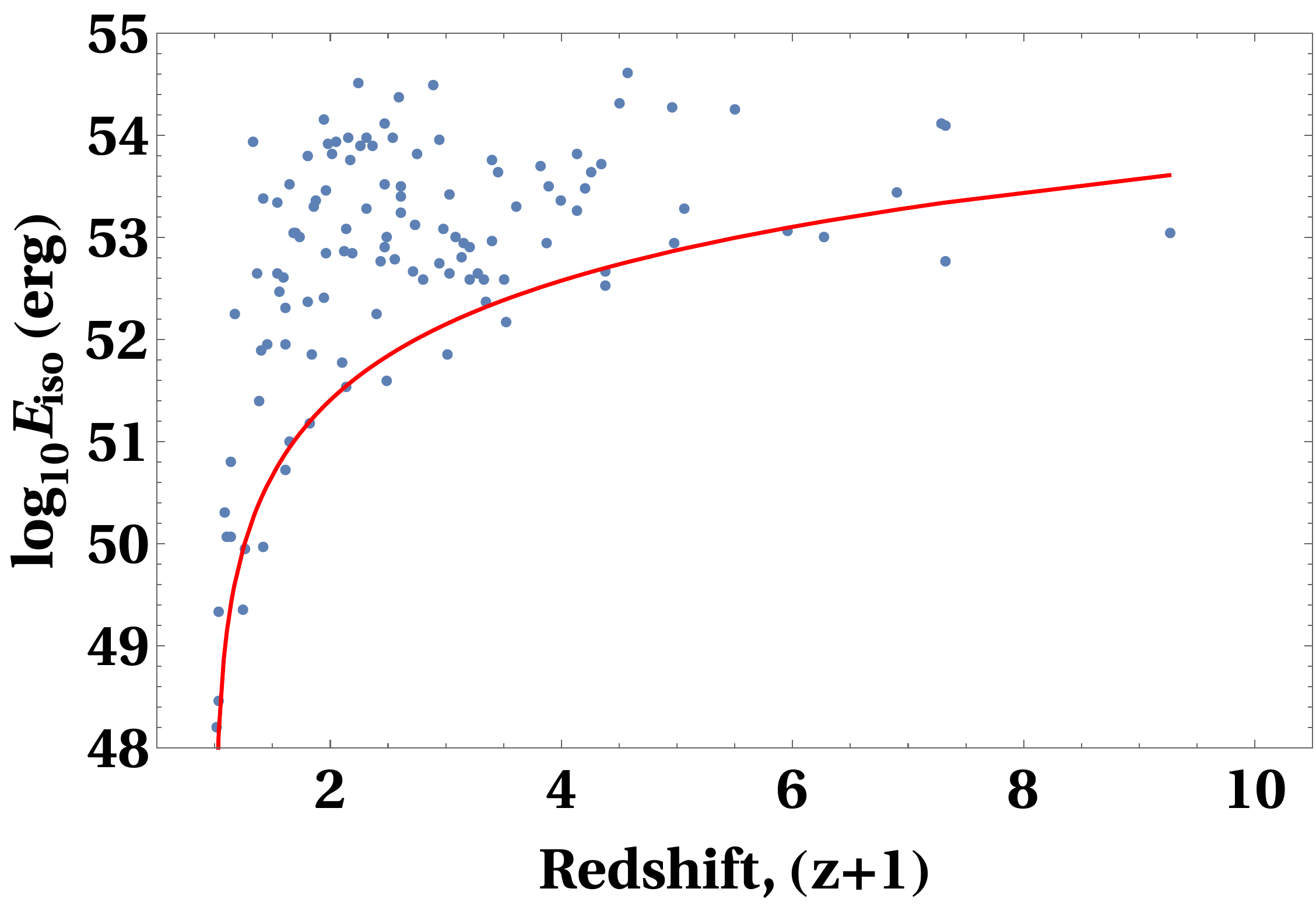}
 \hspace{-0.5cm}   \includegraphics[width=\linewidth,height=6cm]{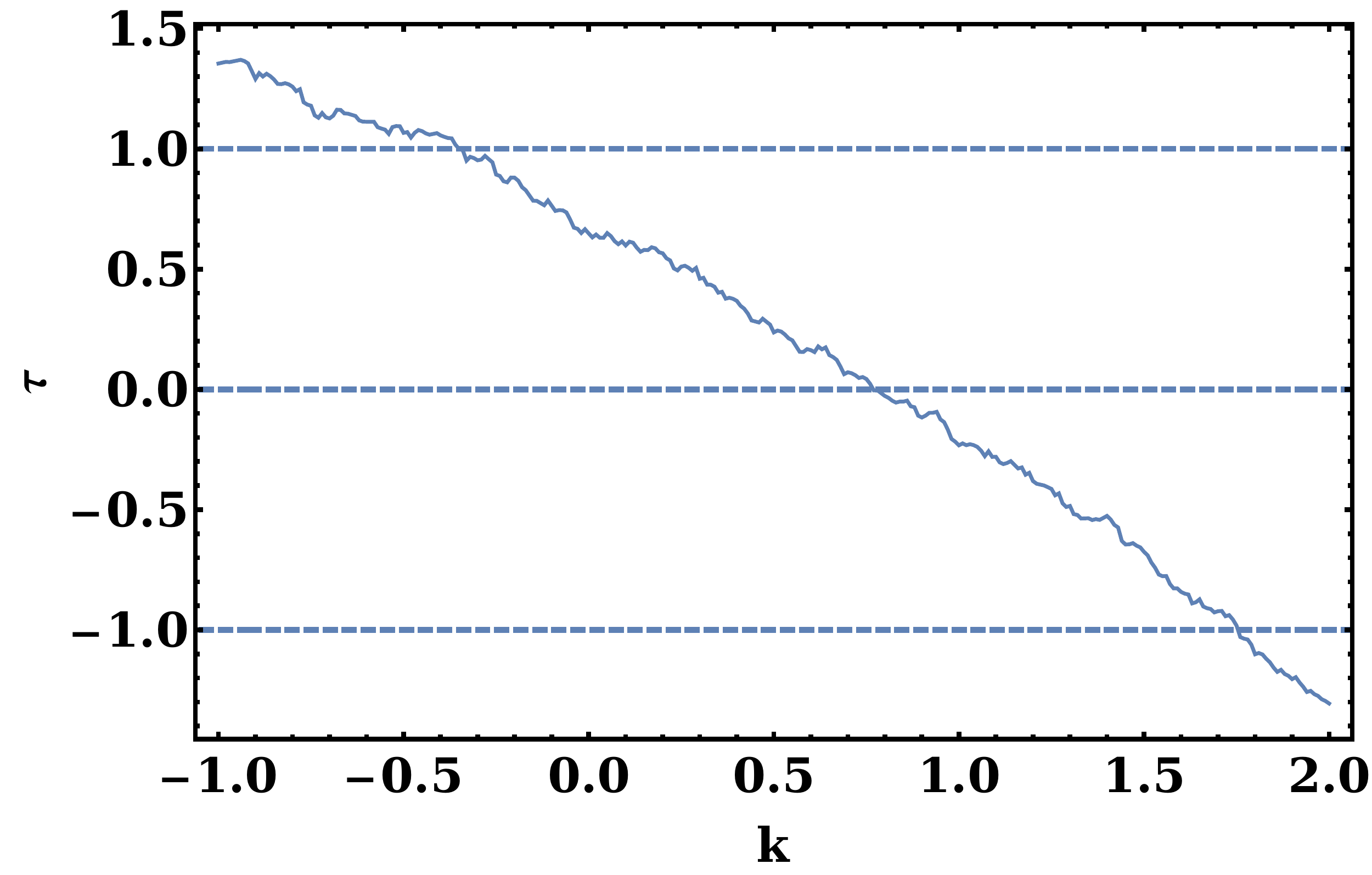}
    \caption{Radio-bright GRB sample:- Upper panel: the $E_{iso}$ vs. redshift distribution with the corresponding limiting fluence = $4.9\cdot10^{-7}$ erg cm$^{-2}$.
    Lower panel: The Kendall tau statistics, $\tau$ vs. k with the blue lines indicating the 1 $\sigma$ confidence level.}
    \label{EP_for_energyRB}
\end{figure}

\begin{figure}
    \includegraphics[width=\linewidth,height=6cm]{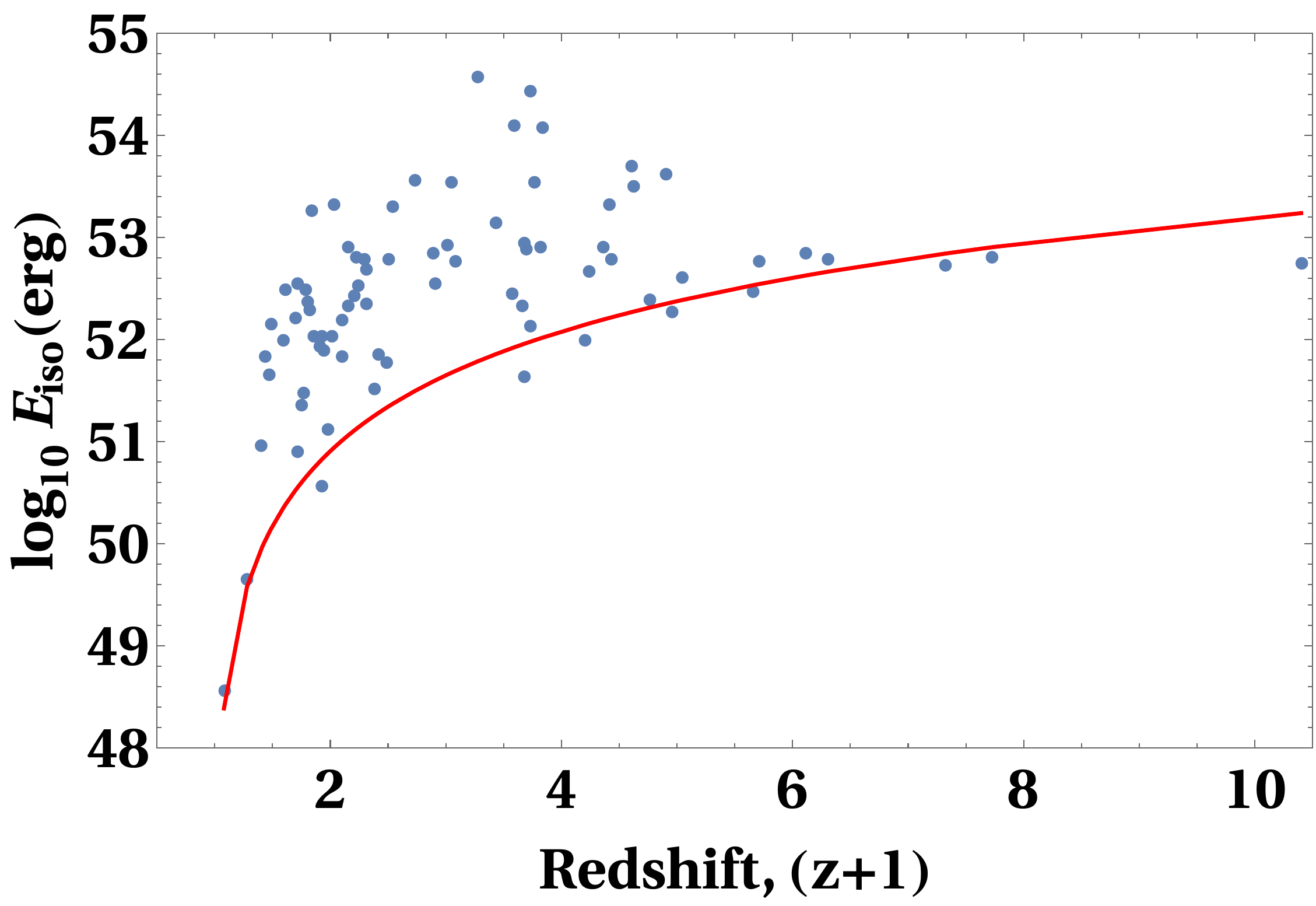}
    \includegraphics[width=\linewidth,height=6cm]{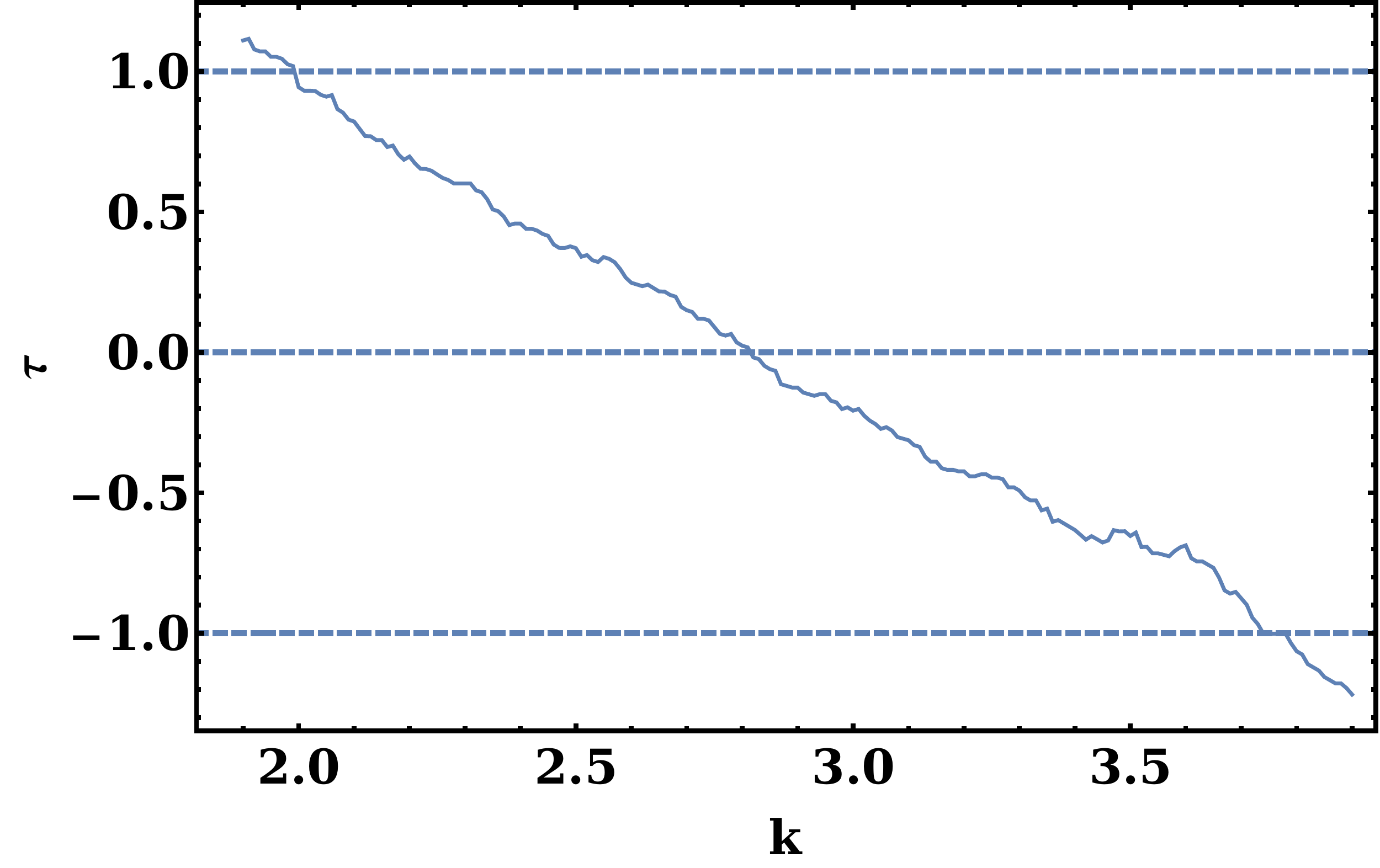}
    \caption{
    Radio-dark GRB sample: Upper panel: the $E_{iso}$ vs. redshift distribution with the corresponding limiting fluence = $1.55\cdot10^{-7}$ erg cm$^{-2}$.
    Lower panel: The Kendall tau statistics, $\tau$ vs. k with the blue lines indicating the 1 $\sigma$ confidence level.}
    \label{EP_for_energyRD}
\end{figure}

We investigate the correlations between different parameters in our radio-bright and radio-dark samples. We show the full matrix in Figure \ref{correlation} but focus on a few significant relevant correlations in this section. Table \ref{tab:coefficients} summarizes the correlation results of our samples, where $k$ indicates the variable's power law dependence on $(1+z)$ (e.g. $T_{int} \propto (1+z)^{k_{T_{int}}})$. Figure \ref{EP_for_energyRB} and Figure \ref{EP_for_energyRD} show the observed isotropic energy, $E_{iso}$ versus $z$ for our sample of 116 radio-bright and 77 radio-dark GRBs.

\subsection*{The correlation between $\theta_{j}$ and redshift}
\cite{LR19c} examined the relationship between jet opening angle and redshift in the entire GRB sample (not separating out radio-dark and bright sub-samples) and found an anti-correlation $\theta_{j} \propto (1 + z)^{−0.75 \pm 0.2}$, suggesting that GRBs at higher redshifts are more tightly collimated than those at low redshifts. Similar trends are reported in \cite{2018LB}.  \cite{LR19c} explain  the correlation quantitatively through an evolving top-heavy IMF, where denser, lower metallicity stars at high redshift can collimate the jet more effectively. 

\begin{figure}
   \hspace{-0.5cm} \includegraphics[width=0.5\textwidth,height=6cm]{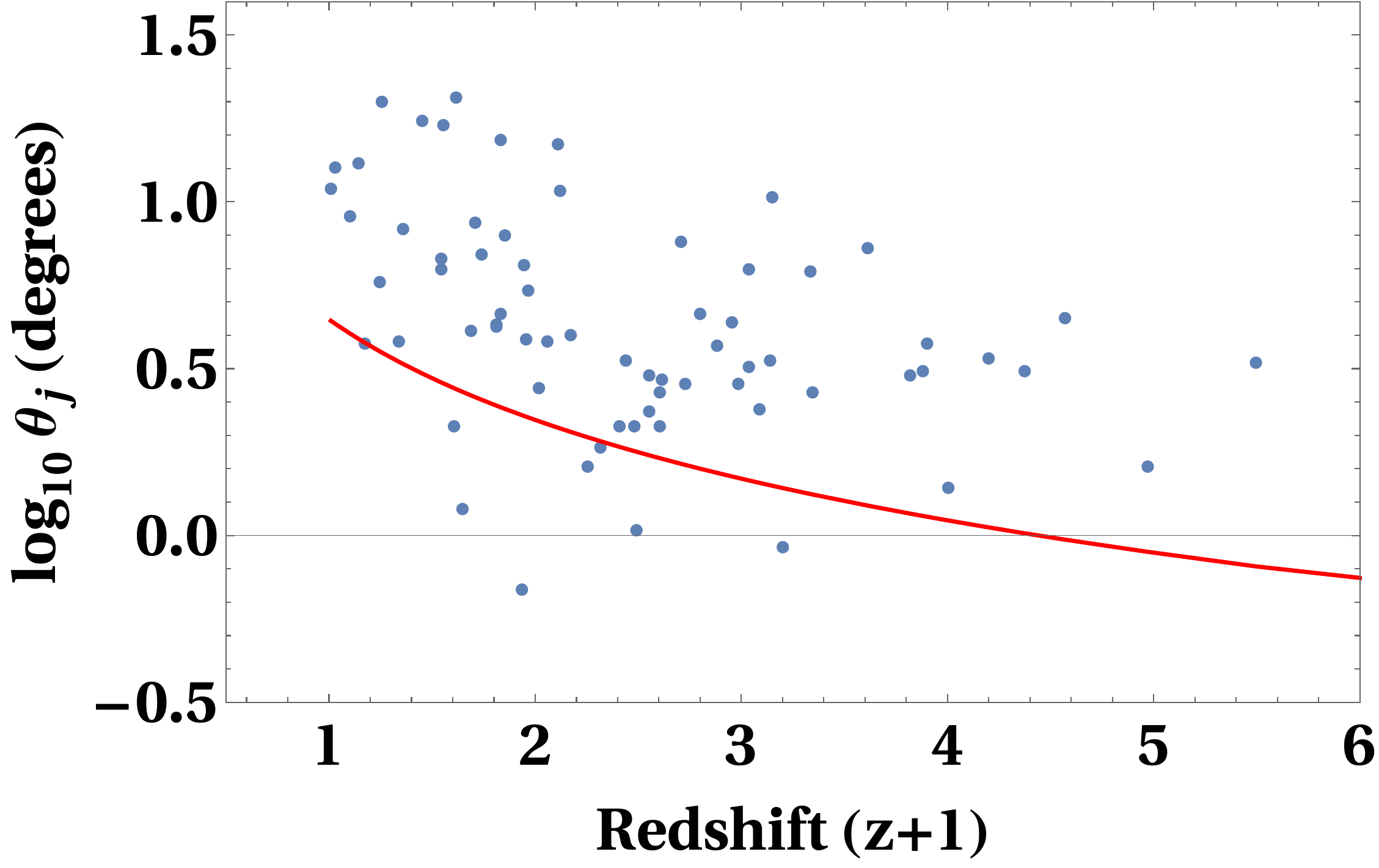}
   \hspace{-0.5cm} \includegraphics[width=0.5\textwidth,height=6cm]{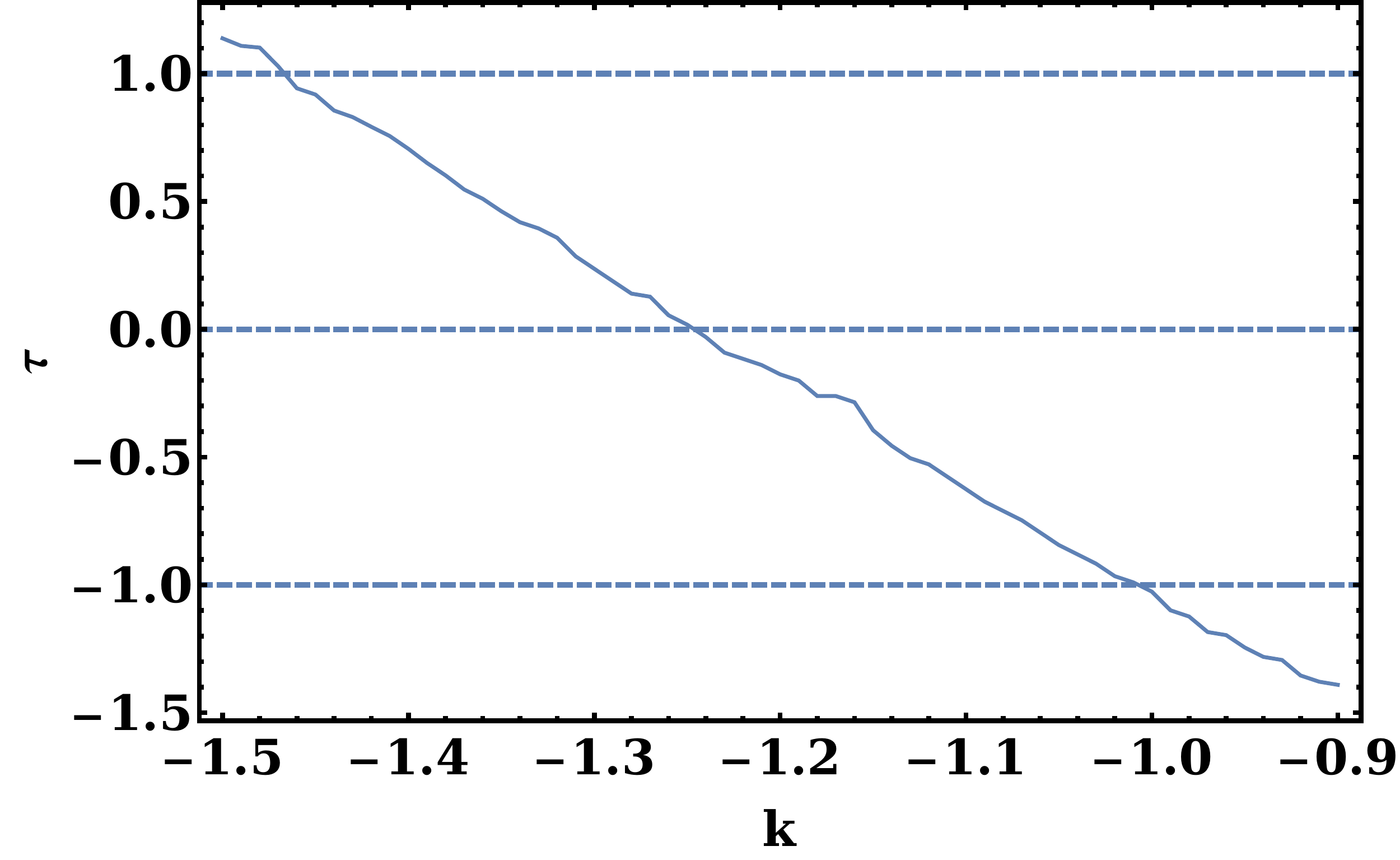}
    \caption{
    Radio-bright GRB sample: Upper panel: the $\theta_{j}$ vs. redshift distribution with the corresponding limiting jet opening angle = 0.69/(1+$z$) degree.
    Lower panel: The Kendall tau statistics vs. k with the blue lines indicating the 1 $\sigma$ confidence level.}
    \label{thetascatterRB}
\end{figure}

\begin{figure} 
    \includegraphics[width=0.5\textwidth,height=6cm]{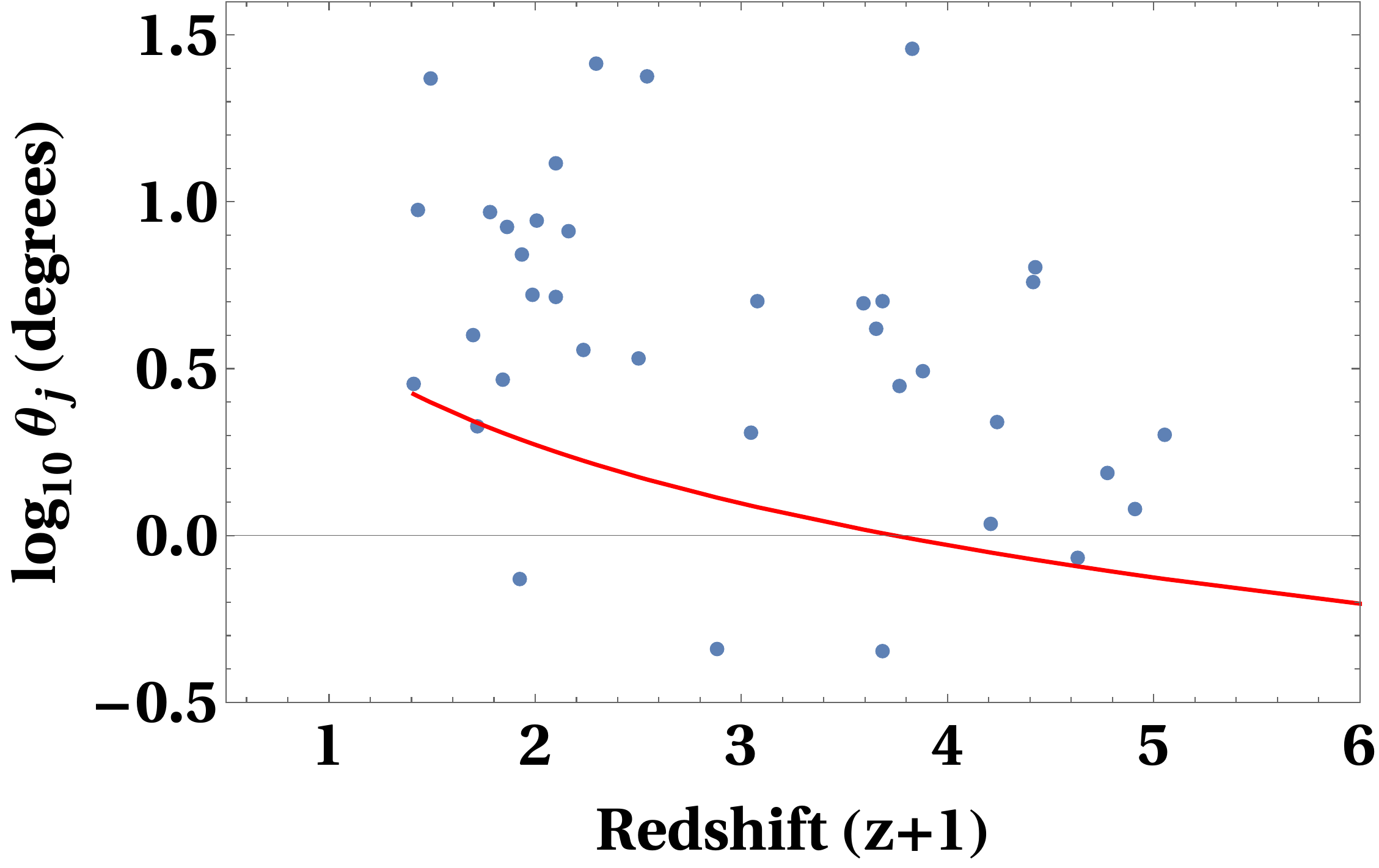}
    \includegraphics[width=0.5\textwidth,height=6cm]{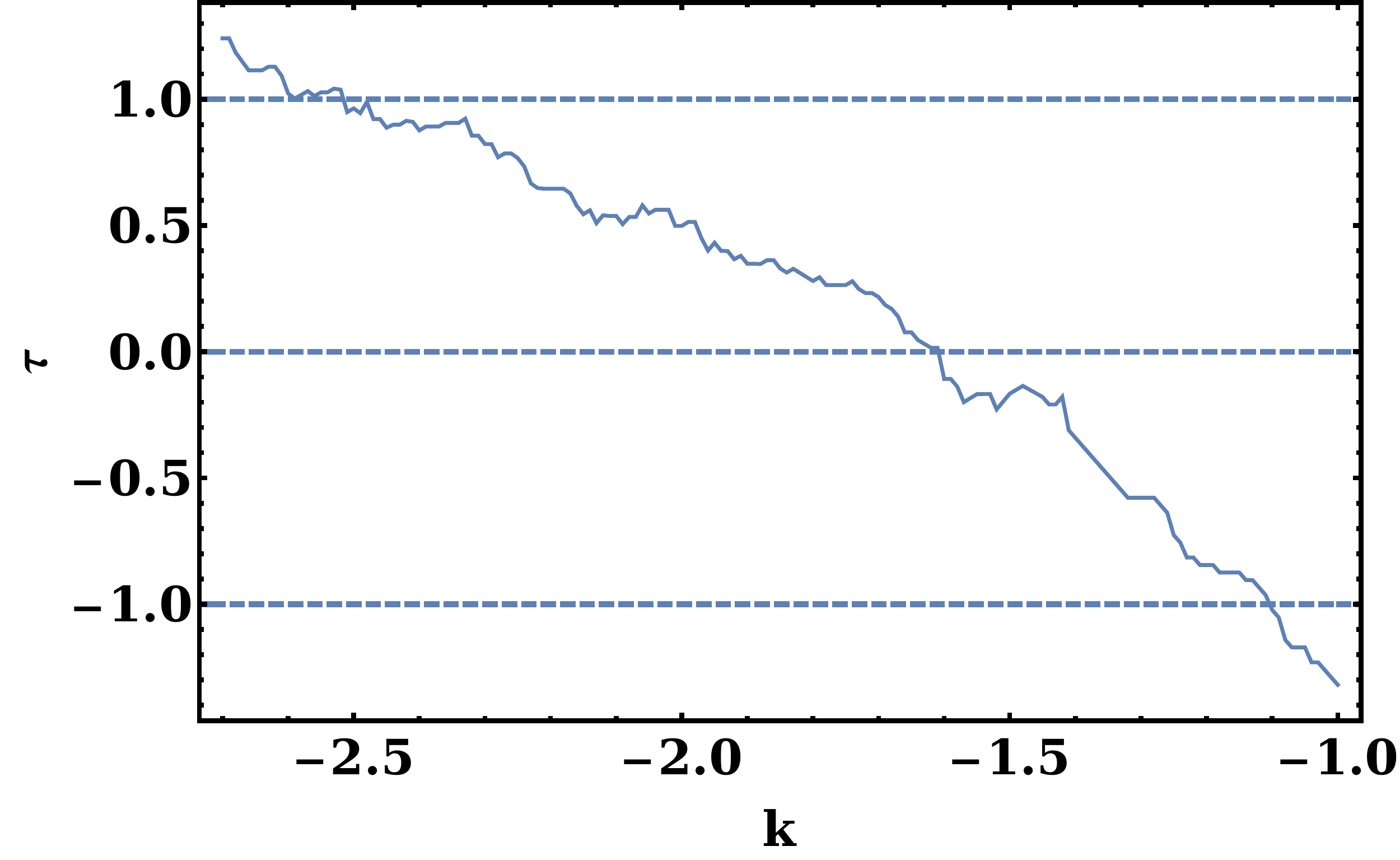}
    \caption{Radio-dark GRB sample: Upper panel: the $\theta_{j}$ vs. redshift distribution with the corresponding limiting jet opening angle = 0.45/(1+$z$) degree.
    Lower panel: The Kendall tau statistics vs. k with the blue lines indicating the 1 $\sigma$ confidence level.}
    \label{thetaRD1}
\end{figure}

We find in our sample a strong anti-correlation is seen for both radio-bright and radio-dark GRBs. It is  parametrised as $\theta_{j} \propto (1 + z)^{-1.3{^{-0.2}_{0.3}}}$ and $\theta_{j} \propto (1 + z)^{−1.6{^{-0.9}_{0.5}}}$ for radio-bright and radio-dark GRB samples, respectively.\footnote{ We note that our Kendell's $\tau$ test sometimes has asymmetric error bars.  For example, we find that for the radio-bright sample, $\theta_{j} \propto (1+z)^{-1.3{^{-0.2}_{0.3}}}$. We report asymmetric error bars, and we have also rounded our power law index to the nearest tenth.}
Therefore, within the error bars, we find the same quantitative dependence of jet opening angle with redshift as \cite{LR19c} in both samples. Figures \ref{thetascatterRB} and Figure \ref{thetaRD1} show the jet opening angle, $\theta_{j}$ versus redshift for our sample.


\subsection*{The correlation between $T_{int}$ and redshift} \label{timez}

In their analysis of radio-bright and dark GRBs with isotropic energies above $10^{52}$ erg, \cite{LR19}  found $T_{int} \propto (1 + z)^{-1.4 \pm 0.3}$ for radio-bright GRB sample and $T_{int} \propto (1 + z)^{-0.4 \pm 0.5}$ for radio-dark GRB sample.  

\begin{figure}
    \includegraphics[width=0.5\textwidth,height=6cm]{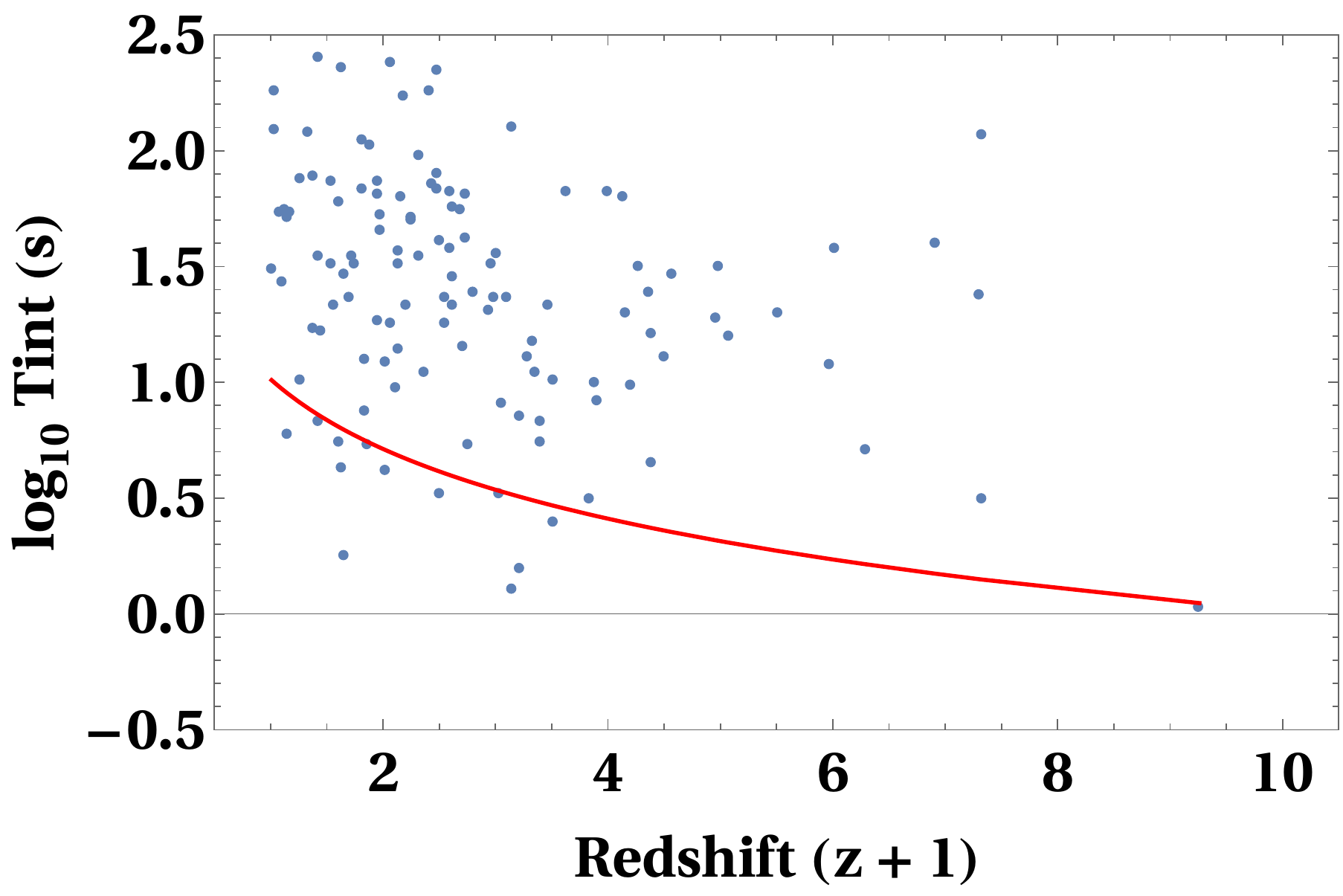}
    \includegraphics[width=0.5\textwidth,height=6cm]{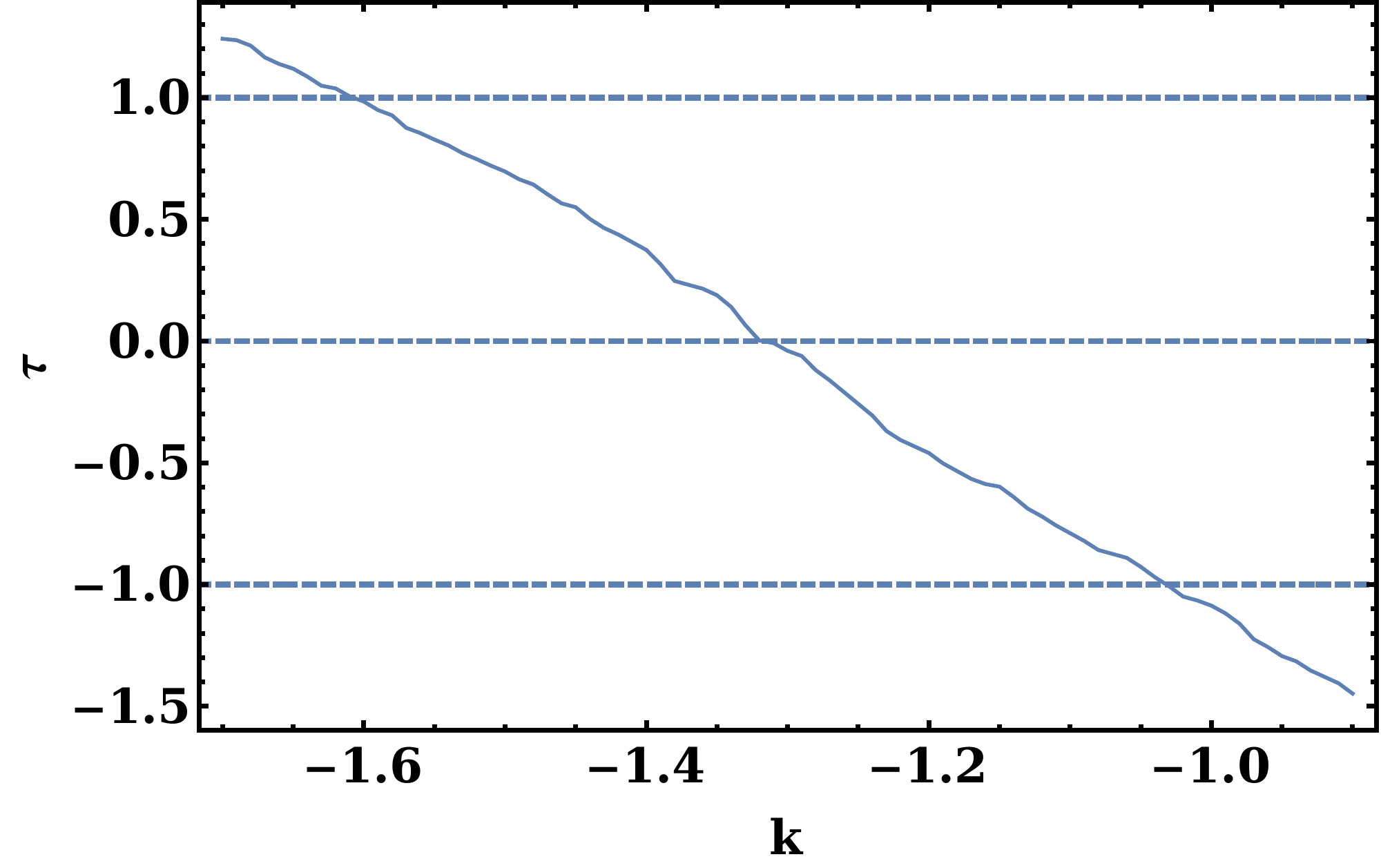}
    \caption{radio-bright GRB sample: Upper panel: $T_{int}$ vs. redshift distribution with the corresponding limiting time = 1.08/(1+$z$) s.
    Lower panel: The Kendall tau statistics, $\tau$, vs. k with the blue lines indicating the 1 $\sigma$ confidence level.} 
    \label{timeRB1}
\end{figure}

\begin{figure}
    \includegraphics[width=0.5\textwidth,height=6cm]{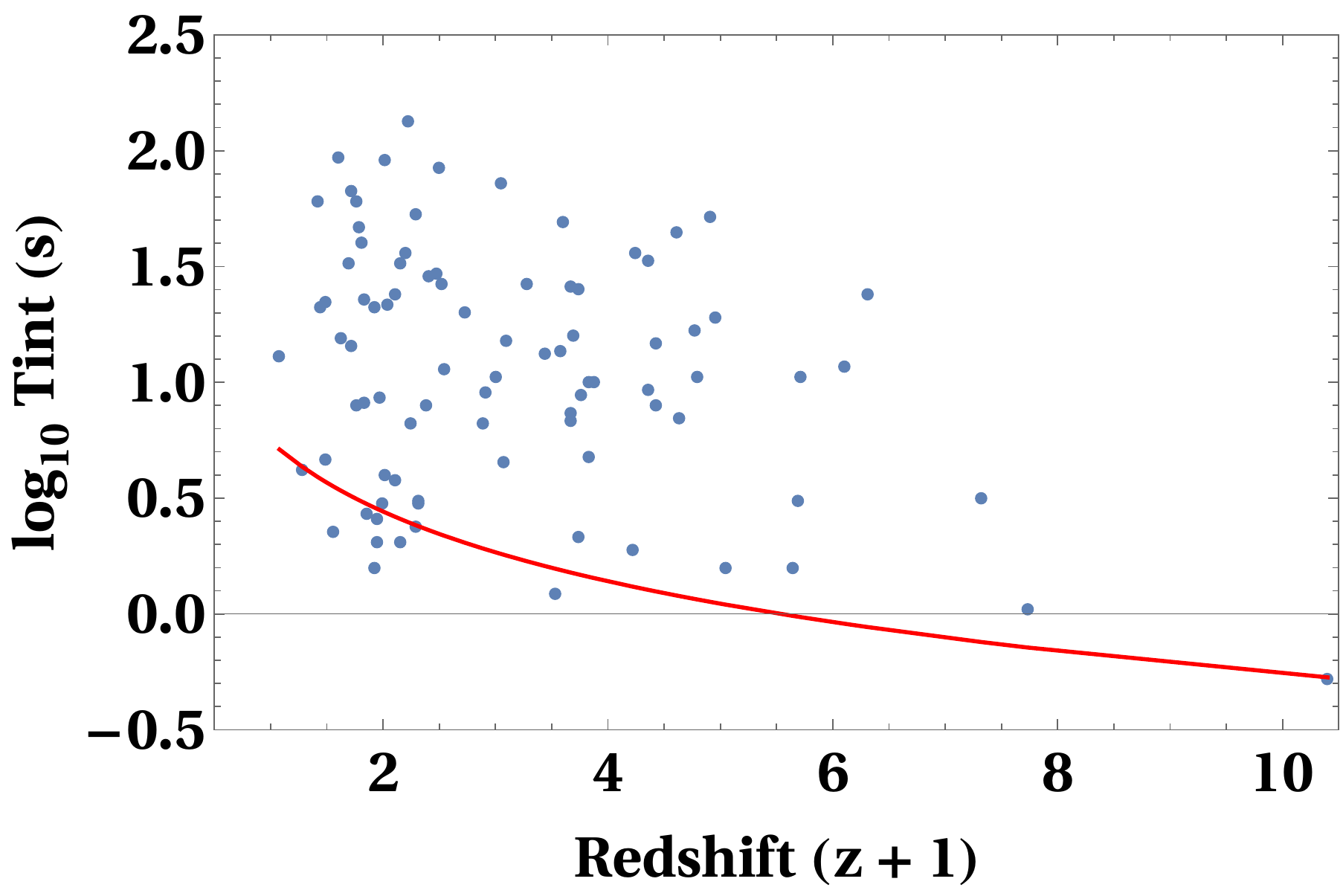}
    \includegraphics[width=0.5\textwidth,height=6cm]{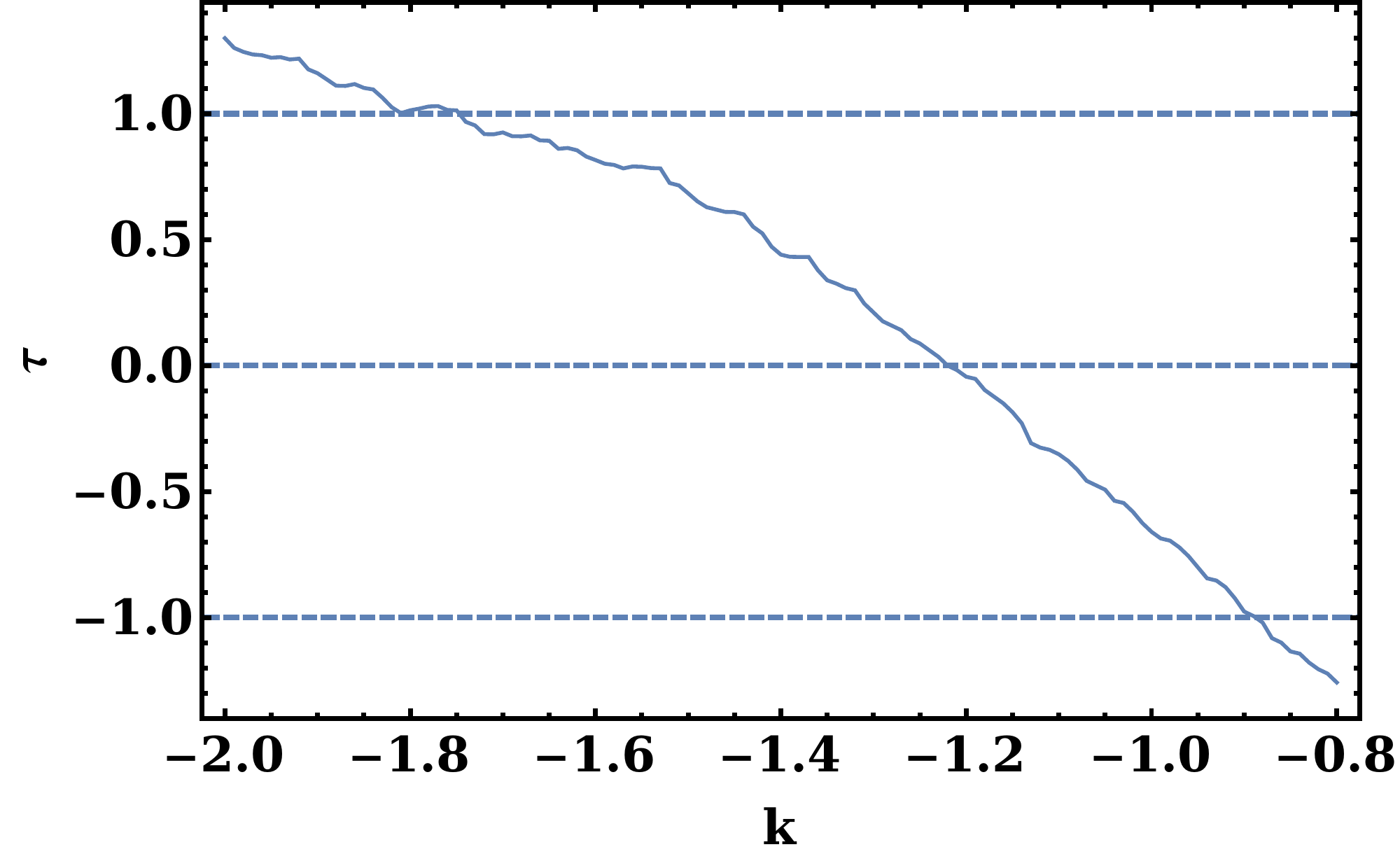}
    \caption{radio-dark GRB sample: Upper panel:  $T_{int}$ vs. redshift distribution with the corresponding limiting time = 0.53/(1+$z$) s.
    Lower panel: The Kendall tau statistics, $\tau$, vs. k with the blue lines indicating the 1 $\sigma$ confidence level.}
    \label{timeRD1}
\end{figure}

Figure \ref{timeRB1} and Figure \ref{timeRD1} show the  $T_{int}$ versus redshift for our sample of 122 radio-bright and 88 radio-dark GRBs and the evolution of Kendall $\tau$. The correlation can be parameterized as $T_{int} \propto (1 + z)^{-1.3{^{-0.3}_{0.3}}}$ for the radio-bright GRB sample and $T_{int} \propto (1 + z)^{-1.2{^{0.3}_{-0.6}}}$ for the radio-dark GRB sample. 
Our values for the radio-bright GRB and radio-dark GRB samples agree with previously reported values in \cite{LR19} within the 1 $\sigma$ error bars. 
Our values are also compatible with the value ($−0.65 \pm 0.27$) reported in \cite{2021DL} within 2 $\sigma$ for both GRB samples.  



\subsection*{Energy-duration correlation}
\begin{figure}
    \includegraphics[width=0.5\textwidth,height=6cm]{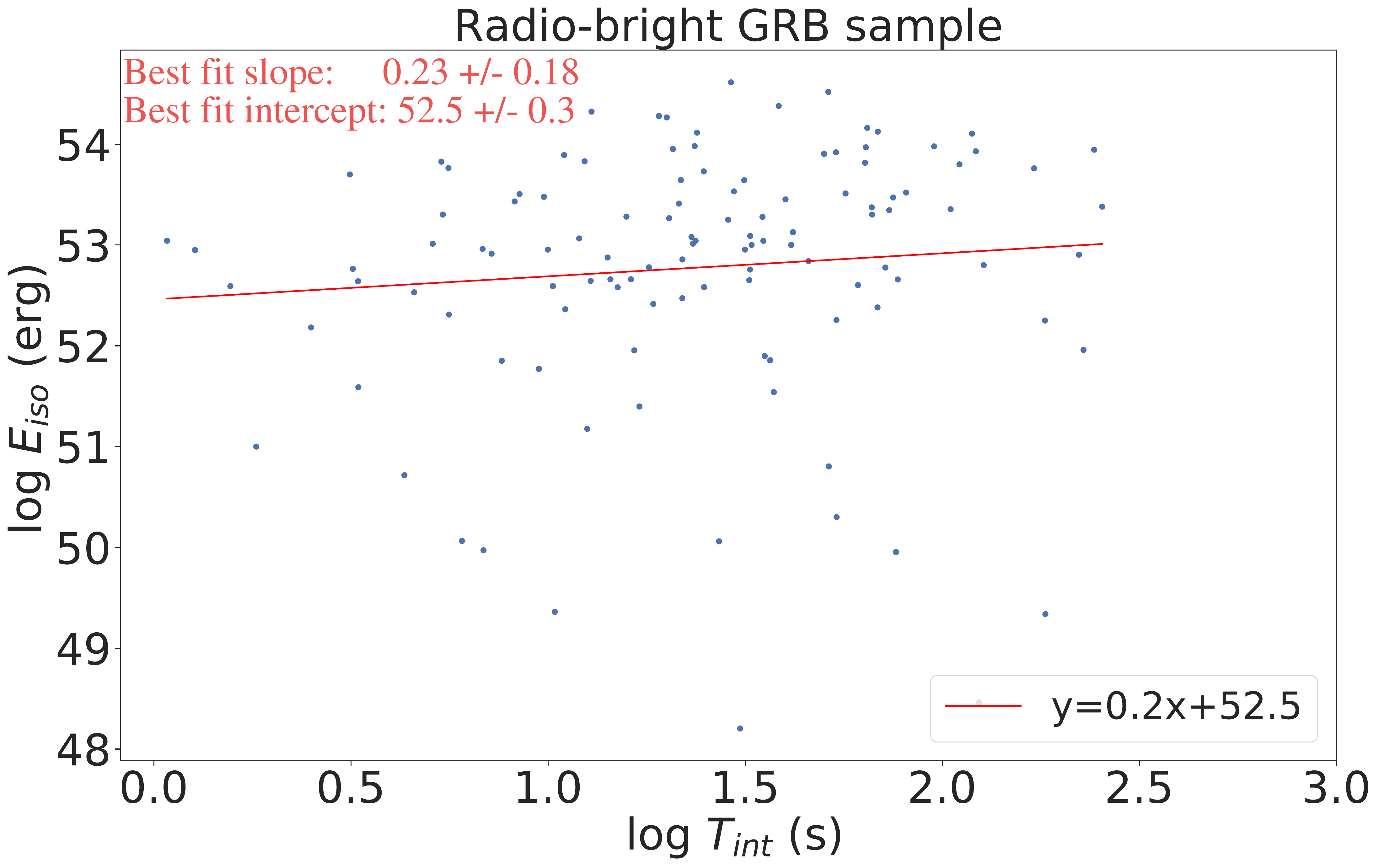}
    \includegraphics[width=0.5\textwidth,height=6cm]{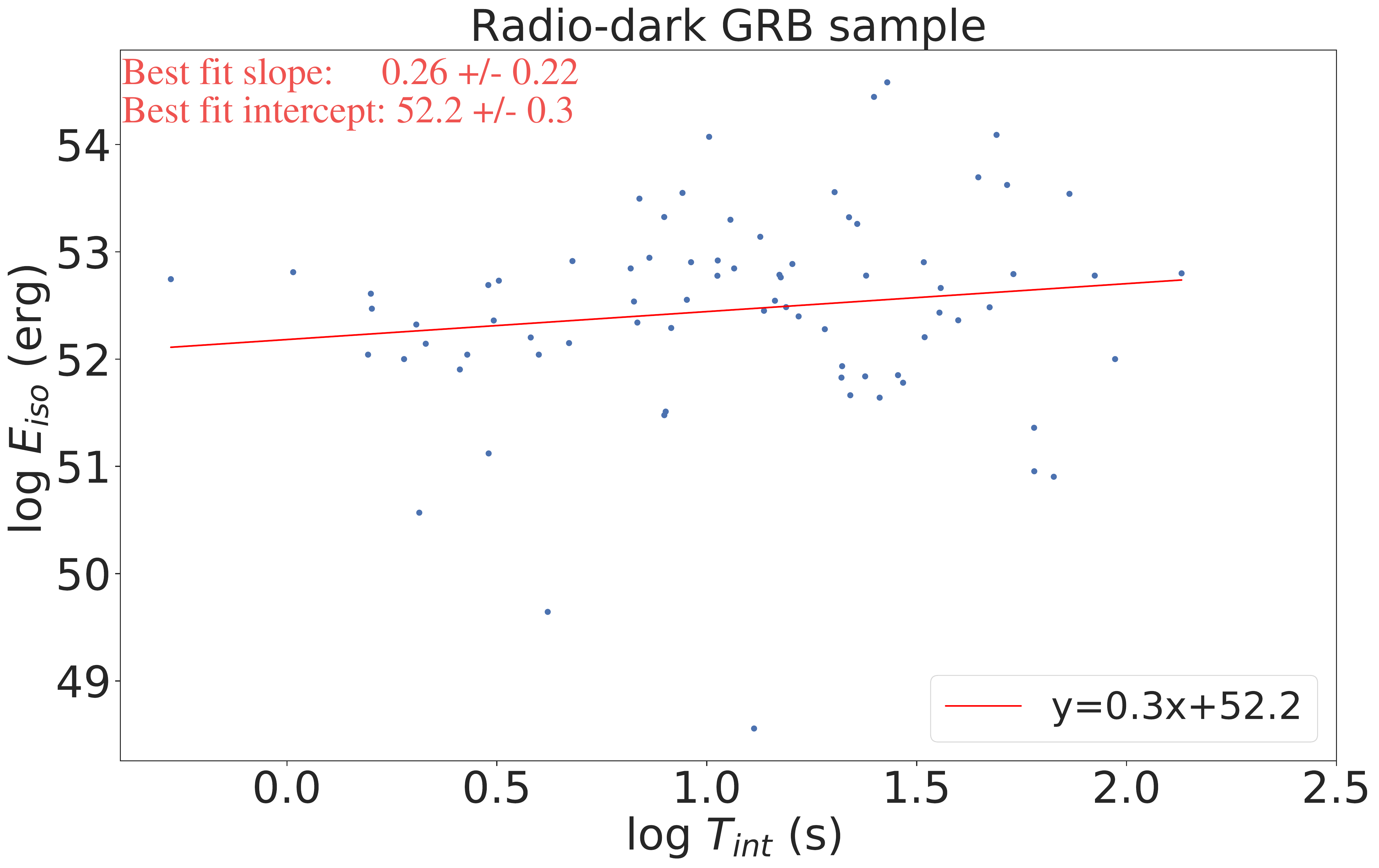}
    \caption{Uncorrected $E_{iso} - T_{int}$ for radio-bright (upper panel) and radio-dark GRB sample (lower panel). In red, we indicate the best-fitting line using the least squares method.}
    \label{en_t}
\end{figure}

\begin{figure}
    \includegraphics[width=0.45\textwidth,height=6cm]{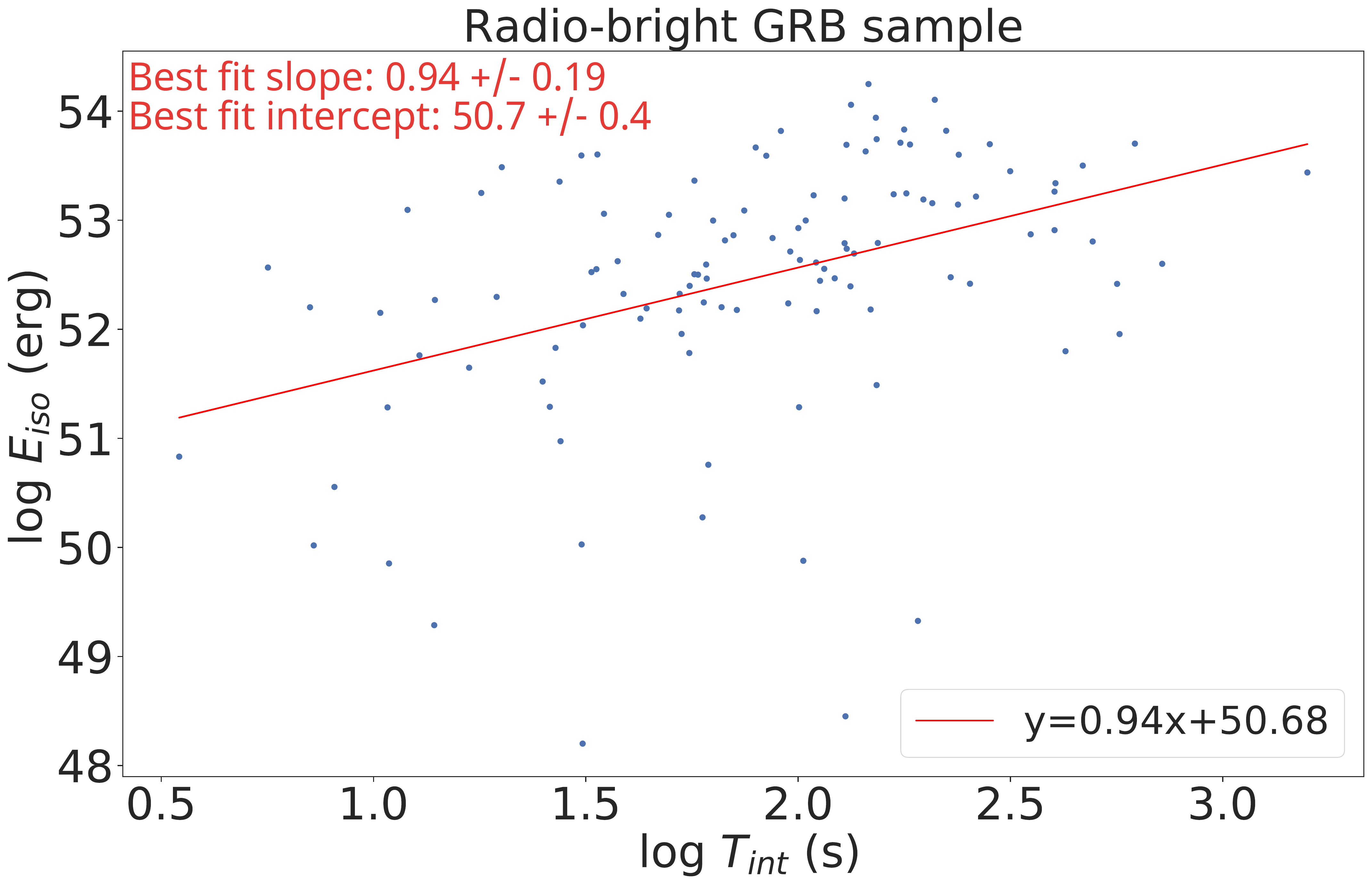}
    \includegraphics[width=0.44\textwidth,height=6cm]{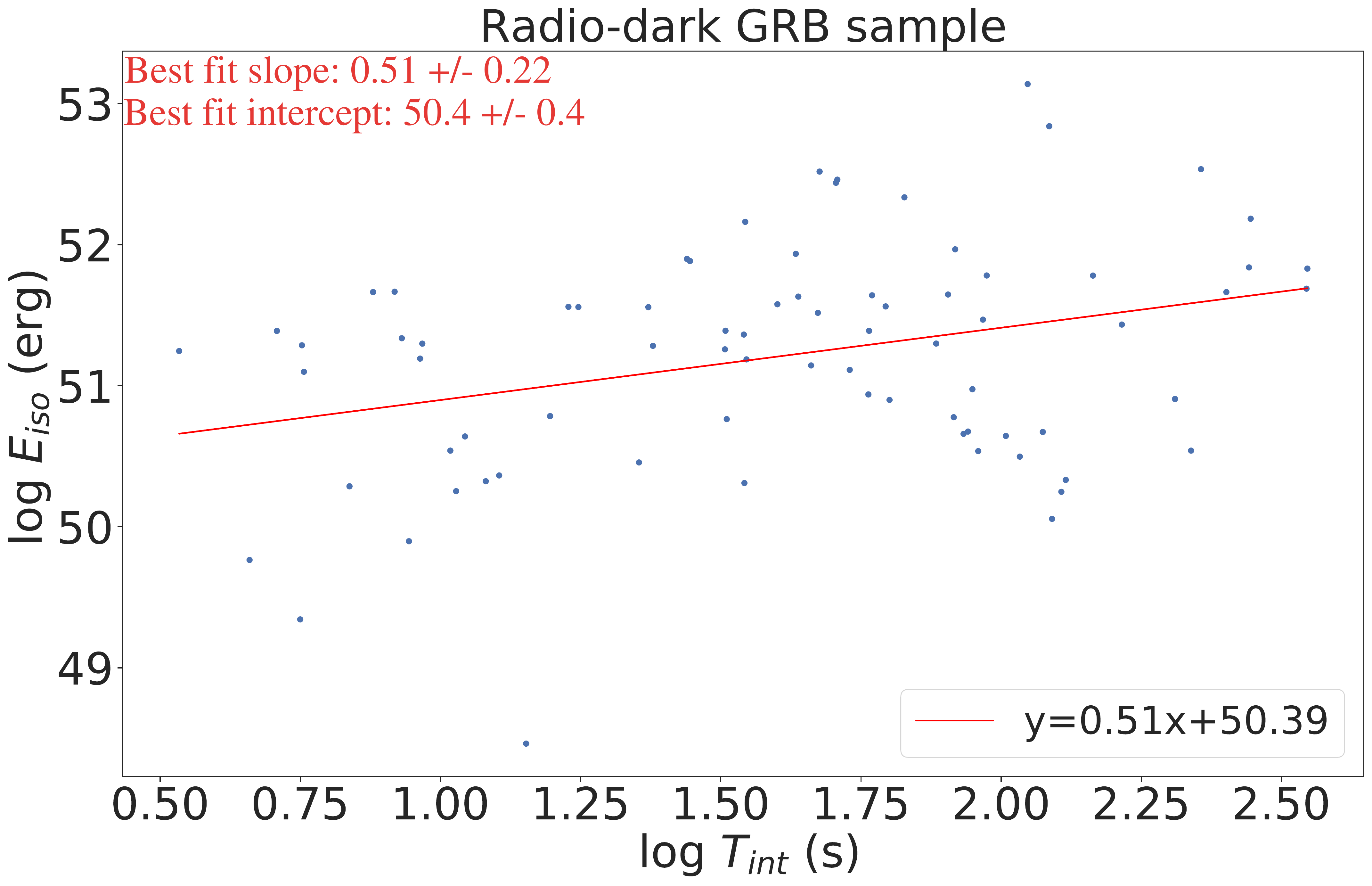}
    \caption{The corrected $E_{iso} - T_{int}$ for radio-bright (upper panel) and radio-dark GRB sample (lower panel). The redshift dependencies have been removed. In red, we indicate the best-fitting line using the least squares method.}
    \label{EPen_t}
\end{figure}

We must remove each variable's dependence on redshift when investigating the correlation between $E_{iso}$ and $T_{int}$. We note \cite{LR19c} found only a weak correlation between these variables and only in their radio-dark sub-sample. 
After correcting for this so that the variables are independent, we find that the $E_{iso}$ is positively correlated with $T_{int}$ for both samples, to high statistical  significance ($> 5 \sigma$). For the radio-bright sample, we find $E_{iso} \propto T_{int}^{0.9 \pm 0.2}$ (see Figure \ref{en_t} and Figure \ref{EPen_t}).  For the radio-dark sample, we find $E_{iso} \propto T_{int}^{0.5 \pm 0.2}$. These results hint that LGRB progenitors will have a longer duration if they have higher isotropic energy at a given redshift. These are also comparable to the values obtained in previous studies \citep{2018T, 2015SN,2019THG}. 


If we consider only radio-bright GRBs with $E_{iso}$ above $10^{52}$ erg and when their redshift dependence is removed, the Pearson's correlation coefficient for $E_{iso}$ vs. $T_{int}$ is much weaker (only $\sim 2 \sigma$ significance). However, for the radio-dark sample, in this case, we obtain a statistically significant correlation $> 4 \sigma$. 


\subsection{Very High Energy Emission}
The Fermi-LAT mission has successfully detected very high-energy $\gamma$-ray emissions (VHE) from GRBs, i.e. emissions in the 0.1 to 100 GeV range. In \cite{2019B}, 186 GRBs (17 short and 169 long) were reported showing VHE by Fermi-LAT. We examined our sample for VHE. Using the data from \cite{2018N}, we found 11 GRBs in the radio-bright sample with very high energy extended emission. There were none found in the radio-dark sample so far. This is consistent with what was found in \cite{LR19}, when they examined the presence of VHE in their sample, and agrees with recent results by \cite{2022LD}. The properties of GRBs with VHE are reported in Table \ref{tab:vhe}. These results suggest a difference in the progenitor or environment of the two GRB subsamples.

The origins of VHE are still unclear. \cite{2019A} suggested a combination of synchrotron and synchrotron self-Compton emission (SSC) as the dominant contributing mechanism toward the VHE spectrum.  \cite{2021ZMY} showed that a combination of the external inverse-Compton and SSC mechanism in the presence of long-lasting central engine activities could explain VHE emission. The bulk Lorentz factor plays an important role, and higher external density environments are preferred, so the presence of this emission can help us constrain both the central engine and the progenitor environment.    However, \cite{2021H} shows that too dense of a medium (e.g., a wind medium) cannot produce the energy seen in the VHE emission. They suggested VHE is more likely to be present in an interstellar medium. Studying VHE emission will help shed light on the particle acceleration mechanism, GRB afterglow phenomena and even the central engine activities, and future detections will help us draw more robust statistical conclusions.



\section{Conclusion and Discussion}  \label{discussion}
In this study, we carried out a systematic investigation of a sample of 123 radio-bright GRBs and 88 radio-dark GRBs with the purpose {of further examining the observational differences between the two sub-populations, as well as testing the results of \cite{LR17} and \cite{LR19}. We use the Student's t-test and KS test to quantify the difference between distributions of variables in our samples and the powerful EP method to obtain an intrinsic correlation between certain variables, especially $T_{int}$ and $\theta_{j}$ with $z$, and $E_{iso}$ with $T_{int}$.  In addition to increasing the sample size, we also, in contrast to \cite{LR17, LR19} which analyzed a sub-set of bright GRBs (those with isotropic energy $E_{iso} > 10^{52} erg$ to mitigate selection effects), we look at the broader sample without this energy cut. We also examine the cosmological evolution of - and correlation between - certain variables beyond what was examined in these previous studies.  Overall, we agree with previous results - that the prompt $\gamma$-ray duration and isotropic energy are larger in radio-bright GRBs compared to the radio-dark population. But we also find some interesting differences worth further exploration. A summary of the main results of the paper, and the differences between previous studies, is:} \\



\begin{itemize}

    \item In agreement with previous results but with a larger sample size, the distributions of $T_{int}$ and $E_{iso}$ are significantly different between the radio-bright and radio-dark sub-populations, with significantly smaller mean values of $T_{int}$ and $E_{iso}$ for the radio-dark sample compared to the radio-bright sample. Both samples have a similar redshift distribution, however. This appears to hold for GRBs with isotropic energies above $10^{52} erg$ and for the whole sample. The differences in these distributions strongly suggest that these samples originated from diverse central engines.  \\


    \item There is a statistically significant anti-correlation between the jet opening angle and redshift, with $\theta_{j} \propto (1 + z)^{-1.3{^{-0.2}_{0.3}}}$ for radio-bright and $\theta_{j} \propto (1 + z)^{−1.6{^{-0.9}_{0.5}}}$ for radio-dark GRB samples. This correlation was seen in the wider GRB sample (without delineating between radio-bright and radio-dark sub-samples in \cite{LR19c}, with a functional form of $\theta_{j} \propto (1+z)^{-0.75 \pm 0.25}$.  We note that, within the error bars, these relations are all consistent. However, the difference in the power-law index is worth future examination as we obtain more jet opening angle measurements and can increase the statistics of this analysis. \cite{LR20} suggested such a correlation can arise from an evolving IMF, with a top-heavy IMF at high redshift.  These higher-mass stars have density profiles that can more effectively collimate the GRB jet as it punches through the stellar envelope. Hence, this relationship can potentially help provide clues to the progenitor systems of these objects. \\

    \item We find an anti-correlation between $T_{int}$ and redshift for both
    radio-bright and radio-dark samples, with the correlation parameterized as $T_{int} \propto (1 + z)^{-1.3{^{-0.3}_{0.3}}}$ and $T_{int} \propto (1 + z)^{-1.2{^{0.3}_{-0.6}}}$, respectively. \cite{LR19c} found a statistically significant anti-correlation between $T_{int}$ and redshift among the whole sample of GRBs with redshifts, with a function form of $T_{int} \propto (1+z)^{-0.8 \pm 0.3}$, while \cite{LR19c} found this correlation only exists in their radio-bright sub-sample (of the brightest GRBs with $E_{iso} > 10^{52} erg$), with a functional form of $T_{int} \propto (1 + z)^{-1.4 \pm 0.3}$. They did {\em not} find a statistically significant anti-correlation between intrinsic duration and redshift in their radio-dark sub-sample, contrasting our results.  This may be due to their much smaller sample size, their selection of only those GRBs with $E_{iso} > 10^{52} ergs$, and the stringent way they accounted for selection effects in the $T-(1+z)$ plane. As the sample size increases, we may continue to test our results' robustness, which suggests the anti-correlation is indeed there in both radio-bright and radio-dark samples. \\

    \item A positive correlation exists between $E_{iso}$ and $T_{int}$ for both radio-bright and radio-dark GRB samples, showing that GRBs with larger isotropic energy will have longer intrinsic prompt duration. For, the radio-bright sample, we find $E_{iso} \propto T_{int}^{0.9 \pm 0.2}$, while for the radio-dark sample, we find $E_{iso} \propto T_{int}^{0.5 \pm 0.2}$. However, \cite{LR19c} found only a very tenuous correlation $\sim 2.5 \sigma$ between $E_{iso}$ and $T_{int}$ in their radio-dark sub-samples.  The difference between our results and theirs may have to do with our increased sample size.\\

     \item In agreement with previous results, we find very high energy $\gamma$-ray emission in the extended emission in the radio-bright GRB sample only, which may further support the hypothesis that radio-bright and radio-dark samples originate from different sources/environments.

\end{itemize}

It is important to note once again that understanding the selection effects at play (and any potential contamination of our sample) is key before drawing any physical conclusions from our results.  In particular, we note that in Table 2 (our radio dark sample), some of the ``radio luminosities'' (which fall below the detector $3 \sigma$ limits) are comparable to the radio bright luminosities.  As such, GRBs that fall along these sensitivity limits can potentially contaminate the sample (i.e. "radio bright" GRBs would be mistakenly classified as "radio dark").  This type of contamination is particularly well-addressed by the Efron-Petrosian method (although we acknowledge there are other ways to characterize and address sensitivity limits; however, these often involving assumed parameterizations of the underlying population distributions). We have tried several different sensitivity parameterizations and truncations, confirming the robustness of our results (if anything, the GRBs which lie along the sensitivity limits dilute our results and our conclusions are even stronger when these are removed).  Nonetheless, it is crucial to continue to question and evaluate selection criteria before assuming any correlation is physical.\\

 Under the assumption that these results are reflective of the underlying physics of the progenitor systems, there are important implications. As discussed in the introduction, \cite{LR22} recently suggested that radio-bright GRBs may originate from massive stars collapsing in interacting binary systems, while these systems can provide both the angular momentum and necessary circumburst environment to explain longer duration radio-bright GRBs. Their paper provides analytic estimates in support of this picture, but more detailed numerical simulations are necessary to test this model (Luu et al., in prep).  The rates of occurrences of such systems may help shed light on this hypothesis.  Methods like binary population synthesis have been used to study the evolution of binary systems from the zero-age main sequence (ZAMS) until their final states. \cite{2020CS} used binary population synthesis to reproduce the LGRB rate considering the two channels, chemically homogeneous evolution, and tidal interactions in the binaries. A mixture of channels is strongly preferred over any single channel since over one channel contributes to at least 10$\%$ of the detected population of BBH mergers \citep{2021ZB}. Of all channels, common envelope (CE) and chemically homogeneous evolution (CHE) contribute to over 70$\%$ of the interacting binary systems \citep{2022B}. They argue that the sub-population of BBH mergers with nonzero spin can explain the observed rates of the entire population of luminous LGRBs. Although unlike \cite{2022B}, \cite{2022AA} detected that only a small fraction of LGRBs are likely associated with BBH mergers. Because of the presence of uncertainties in several key aspects of binary evolution (for example, mass and angular momentum losses during binary interactions and stability of mass transfer (MT) and CE evolution), the binary progenitor hypothesis for radio-bright GRBs warrants further investigation.

It is also interesting to consider the role metallicity evolution plays in the trends we see in our data. 
LGRBs prefer low-metallicity host environments. Low metallicity helps mitigate mass and angular momentum losses, which creates favorable conditions for launching the relativistic jets from the central engine. In a binary context, a common envelope phase is more easily achieved in low metallicity binaries than in higher metallicity ones. The low-metallicity stars also have a large helium core mass compared to the high-metallicity stars. Such systems are most likely associated with the formation of BHs \citep{2010BB}. A detailed analysis of the relationship between binary systems, LGRBs, and metallicity evolution and the contribution of evolutionary channels to radio-bright and radio-dark samples warrants further study.

Finally, exploring other avenues or angles to look at this potential dichotomy could offer promising new insights into the nature of GRB progenitors.  For example, as mentioned in \S 2, most supernovae associations fall within our radio-bright sample. With the small sample size and the ill-defined selection effects in this case, the significance of this is currently inconclusive. As more SN/GRB associations are established (or not), we may test whether this is a statistically significant result (and not simply a selection effect), which would certainly add credence to a physical difference in the underlying progenitor and central engine.  Examining differences in the prompt light curve variability or detailed spectral evolution among the two samples may give a further glimpse into the central engine and environment of GRBs and allow us to better understand whether and when this dichotomy is a result of detector sensitivity/selection effects and when it is a reflection of a true physical difference in the underlying GRB progenitor.


\vspace{-5mm}
\section{Acknowledgements}
We thank the referee for very valuable comments and suggestions that improved this manuscript.  We thank Roseanne Cheng, Ken Luu, Jarrett Johnson and Phoebe Upton-Sanderbeck for interesting discussions related to GRB progenitors. We thank B. De Simone for his helpful help on some of Efron and Petrosian analysis. Los Alamos National Laboratory is operated by Triad National Security, LLC, for the National Nuclear Security Administration of U.S. Department of Energy (Contract No. 89233218CNA000001). LA-UR-22-29846 
\vspace{-5mm}
\section{Data Availability}

The data underlying this article will be shared on reasonable request
to the corresponding author.







\bibliographystyle{mnras}
\bibliography{main} 





\onecolumn 

\setlength\LTleft{-10mm}            
\setlength\LTright{0pt}           

\begin{longtable}{ccccccccccc}
\captionsetup{width=\linewidth}
\caption{List of the radio-bright GRBs reporting different parameters. The columns indicate the GRB name, $T_{90}$ and the intrinsic burst duration, $T_{int}$, the redshift, the equivalent energy, $E_{iso}$, luminosity density or spectral peak luminosity at radio band, L$_{R}$, X-ray luminosity, L$_{X-ray}$, and jet opening angles in radian, $\theta_{j}$. In the last column, we indicate the class of bursts, such as XRR, XRR, or GRB-SN GRBs. The radio afterglows detected by the AMI are at 15.7 GHz taken from \citep{2018Anderson, 2021Z}. All other radio afterglows (including those from VLA, EVLA, and ATCA) are taken from \citep{2012CF, 2021Z} at 8.5 GHz. For the measurement of radio luminosity density, redshift information is needed, so for GRBs with missing redshifts, radio luminosity densities cannot be estimated. For GRBs with missing redshift information, radio luminosity density can't be estimated. References are as follows: [1] \citep{2012CF}, [2] \citep{2020W}, [3] \citep{2021Z}, [4] GCN circulars (\url{https://gcn.gsfc.nasa.gov/gcn/gcn3_archive.html}), [5] Greiner's GRB list (\url{https://www.mpe.mpg.de/~jcg/grbgen.html})}  
\label{tab:radiobright}\\ 
\hline
GRB     & $T_{90}$  &  $T_{int}$  & $z$      & $E_{iso}$   & L$_{R}$    & L$_{X-ray}$      & $\theta_{j}$ &   Radio Telescope & Notes & Reference \\ 
        & (s)     & (s)    &        & (erg)     & (erg/s/Hz) & (erg/s) & (rad)    &          &           &           \\ \hline \noalign{\smallskip}
\endfirsthead
\multicolumn{9}{c}%
{{\bfseries Table \thetable\ continued from previous page}} \\ \hline
GRB     & $T_{90}$  &  $T_{int}$   & $z$      & $E_{iso}$   & L$_{R}$    & L$_{X-ray}$      & $\theta_{j}$ & Radio Telescope & Notes & Reference \\
        & (s)     & (s)    &        & (erg)     & (erg/s/Hz) & (erg/s) & (rad)    &         &            &           \\ \hline \noalign{\smallskip}
\endhead
\hline
\endfoot
\endlastfoot
970508  & 14      & 7.63   & 0.835  & 7.1E + 051  & 9.4E + 030  & ...      & 0.269    & VLA   & ...              & {[} 1,2,3 {]}        \\
970828  & 147     & 75.08  & 0.958  & 2.96E + 053 & 2.93E + 030 & ...      & 0.068    & VLA   & ...              & {[} 1,2,3 {]}        \\
980329  & 58      & 12.89  & 3.5    & 2.1E + 054  & 4.76E + 031 & ...      & ...      & VLA   & ...              & {[} 1,2,3 {]}        \\
980425  & 31      & 30.72  & 0.009  & 1.6E + 048  & 8.56E + 028 & ...      & 0.192    & ATCA  & SN / GRB         & {[} 1,2,3 {]}        \\
980519  & 30      & ...    & ...    & ...         & 1.95E + 031 & ...      & 0.047    & VLA   & ...              & {[} 1,2,3 {]}        \\
980703  & 90      & 45.78  & 0.966  & 6.9E + 052  & 1.95E + 031 & ...      & 0.095    & VLA   & ...              & {[} 1,2,3 {]}        \\
981226  & 20      & 9.48   & 1.11   & 5.9E + 051  & 4.4E + 030  & ...      & ...      & VLA   & ...              & {[} 1,2,3 {]}        \\
990123  & 100     & 38.46  & 1.6    & 2.39E + 054 & 1.28E + 031 & ...      & 0.047    & VLA   & ...              & {[} 1,2,3 {]}        \\
\hline
\end{longtable}

\begin{longtable}{ccccccccccc}
\captionsetup{width=\linewidth}
\caption{List of the radio-dark GRBs reporting their different parameters. The columns depict the GRB name, $T_{90}$, $T_{int}$, namely the intrinsic duration of the bursts, redshift, the isotropic equivalent energy $E_{iso}$, upper limits of the luminosity density or spectral peak luminosity at radio band, L$_{R}$, X-ray luminosity, L$_{X-ray}$, and jet opening angles, $\theta_{j}$ in radians. In the last column, we indicate the GRB class. The radio afterglows detected by the AMI are at 15.7 GHz taken from \citep{2018Anderson, 2021Z}. All other radio afterglows (including those from VLA, EVLA, and ATCA) are taken from \citep{2012CF, 2021Z} at 8.5 GHz. Due to the paucity of measurement of the radio afterglows with the upper limits, the radio luminosity density values of radio-quiet afterglows can only be estimated as the upper limits.  References are as follows: [1] \citep{2012CF}, [2] \citep{2020W}, [3] \citep{2021Z}, [4] GCN circulars (\url{https://gcn.gsfc.nasa.gov/gcn/gcn3_archive.html}), [5] Greiner's GRB list (\url{https://www.mpe.mpg.de/~jcg/grbgen.html})} 
\label{tab:radiodark} \\
\hline
GRB     & $T_{90}$   & $T_{int}$   & $z$              & $E_{iso}$      &L$_{R}$   & L$_{X-ray}$       & $\theta_{j}$ & Radio Telescope & Notes & Reference \\
        & (s)   & (s)    &                & (erg)     & (erg/s/Hz) & (erg/s) & (rad)    &          &           &           \\ \hline \noalign{\smallskip}
\endfirsthead

\multicolumn{9}{c}%
{{\bfseries Table \thetable\ continued from previous page}} \\ \hline
GRB     & $T_{90}$   & $T_{int}$   & $z$              & $E_{iso}$      &  L$_{R}$   & L$_{X-ray}$       & $\theta_{j}$ & Radio Telescope & Notes & Reference \\
        & (s)   & (s)    &                & (erg)     & (erg/s/Hz) & (erg/s) & (rad)    &         &            &           \\ \hline \noalign{\smallskip}
\endhead
\hline
\endfoot
\endlastfoot

970228  & 56    & 33.04  & 0.695          & 1.6E + 052  & 8.26E + 029 & ...      & 0.0698  & VLA   & SN / GRB & {[} 1,2,3 {]} \\
971214  & 35    & 7.92   & 3.42           & 2.11E + 053 & 1.14E + 031 & ...      & ...     & VLA   & ...      & {[} 1,2,3 {]} \\
980613  & 50    & 23.84  & 1.097          & 6.9E + 051  & 1.78E + 029 & ...      & ...     & VLA   & ...      & {[} 1,2,3 {]} \\
990705  & 42    & 22.83  & 0.84           & 1.82E + 053 & 1.71E + 029 & ...      & 0.051   & ATCA  & ...      & {[} 1,2,3 {]} \\
990712  & 30    & 20.94  & 0.433          & 6.72E + 051 & 2.58E + 029 & ...      & 0.165   & ATCA  & ...      & {[} 1,2,3 {]} \\
000630  & 20    & ...    & ...            & ...         & ...         & ...      & ...     & VLA   & ...      & {[} 1,2,3 {]} \\
020127  & 26    & 8.97   & 1.9            & 3.57E + 052 & ...         & ...      & ...     & VLA   & ...      & {[} 1,2,3 {]} \\
020305  & 247   & ...    & \textless{}2.8 & ...         & ...         & ...      & ...     & VLA   & ...      & {[} 1,2,3 {]} \\
\hline 
\end{longtable}

\twocolumn

\begin{table}
    \caption{The number of LGRBs considered in our sample includes XRF and XRR GRBs and GRBs with SN association.} 
    \label{tab:subclass}    
    \begin{tabular}{cccc}
        \hline
         Category & Total sample & Radio-bright sample & Radio-dark sample \\
         \hline \hline
         LGRB & 344 & 167 & 177 \\
         SN/GRB & 23 & 21 & 2 \\
         
         XRF & 26 & 11 & 15 \\ 
         XRR & 121 & 48 & 73\\ 
         \hline
    \end{tabular}
\end{table}

\begin{table}
    \caption{Mean values of $z$, $T_{int} (s)$, $E_{iso} (erg)$, for our total sample (upper Table), and only those GRBs with $E_{iso} > 10^{52}$ erg (lower Table).}
    \label{tab:meanvalues}
    \begin{tabular}{cccccc}
        \hline
         Mean & $z$ & $T_{int} (s)$ & $E_{iso} (erg)$  \\
         \hline \hline
         
         Radio-bright (123 GRBs) & 1.83 & 44 & $3.92 \cdot 10^{53}$  \\
         Radio-dark (88 GRBs) & 2.16 & 21 & $1.84 \cdot 10^{53}$  \\
         \hline
         Mean for $E_{iso}> 10^{52}$ erg&  &  &  &  &  \\
         \hline \hline
         Radio-bright(94 GRBs) & 2.13 & 44 & $4.83 \cdot 10^{53}$  \\
         Radio-dark (58 GRBs) & 2.51 & 20 & $2.43 \cdot 10^{53}$  \\
         \hline
         \vspace{1mm}
    \end{tabular}
    \vspace{2mm}

\textbf{Note:} From the mean values of radio-bright and radio-dark GRBs parameters, we can see that the intrinsic duration is longer for the former, and the isotropic equivalent energy is also larger for the former. The redshift and jet opening angles of radio-dark GRBs exceed that of radio-bright GRBs.

\end{table}

\begin{table}
    \caption{Statistical test results for the 123 radio-bright and 88 radio-dark GRBs.}
    \label{tab:stat_tests}
    \begin{tabular}{@{\hskip3pt}ccc@{\hskip3pt}cc@{\hskip3pt}c@{\hskip3pt}}
        \hline
         Probability & $z$ & $T_{int} (s)$ & $E_{iso} (erg)$  \\
         \hline \hline
         KS Test & 0.07 & $5.6 \cdot 10^{-4}$ & $8 \cdot 10^{-6}$  \\
         
         Student's t-test & 0.1 & $2.4 \cdot 10^{-5}$  & 0.03 \\
         \hline
    \end{tabular}
    \vspace{1mm}

\textbf{Note:} The first row shows the KS test p-value. The second row shows the Student's t-test p-value.
\end{table}

\begin{table}
    \caption{Average values and statistical test results for GRB samples with $E_{iso}$ > $10^{52} erg$ (94 radio-bright GRBs and 58 radio-dark GRBs).}
    \label{tab:stat_tests_Eisocut}
%
 %
 \begin{tabular}{@{\hskip3pt}ccc@{\hskip3pt}cc@{\hskip3pt}c@{\hskip3pt}}
        \hline
         Probability & $z$ & $T_{int} (s)$ & $E_{iso} (erg)$     \\
         \hline \hline
         KS & 0.1 & $8.65 \cdot 10^{-4}$ & $3 \cdot 10^{-6}$  \\
         
         Student's t & 0.16 & $1.48 \cdot 10^{-4}$ & 0.04 \\
         \hline
         
    \end{tabular}   \\
\textbf{Note:} We select a subset of GRBs composed of 94 radio-bright and 58 radio-dark GRBs with $E_{iso}$ > $10^{52} erg$.
\end{table}

\setlength{\tabcolsep}{0.5pt}
\begin{table}{
\caption{Average values and statistical test results between samples of GRBs with known jet opening angle measurements.}
\label{tab:stat_tests_jetangles}
    \begin{tabular}{@{\hskip3pt}c|c@{\hskip4pt}c@{\hskip2pt}cc@{\hskip2pt}c@{\hskip2pt}}
        \hline
         Probability & $z$ & $T_{int} (s)$ & $E_{iso} (erg)$ & $E_{\gamma} (erg)$ & $\theta_{j} (deg)$ \\
         \hline \hline
         KS & 0.03 & 0.02 & 0.049 & 0.06 & 0.79 \\
         
         Student's t & 0.13 & $2.35 \cdot 10^{-3}$ & 0.07 & 0.17 & 0.4\\
         \hline
         
         Mean for $E_{iso}> 10^{52}$ erg &  &  &  &  &  \\
         \hline \hline
         Radio-bright & 1.73 & 38.7 & $3.88 \cdot 10^{53}$ & $1.05 \cdot 10^{51}$ & 5.47 \\
         Radio-dark & 2.22 & 19.4 & $1.25 \cdot 10^{53}$ & $5.74 \cdot 10^{51}$ &  6.48 \\
         \vspace{1mm}
    \end{tabular}
    
\textbf{Note:}   All the bursts in this subset have known $\theta_{j}$ and $z$ values. 
}
\end{table}

\begin{table}
\caption{The dependence of $T_{int}$ and $\theta_{j}$ on redshift for both the radio-bright GRBs (first row) and radio-dark GRBs (second row). The parameter $k$ indicates the power-law dependency (e.g. $T_{int} \propto (1+z)^{k_{T_{int}}})$.}

\label{tab:coefficients}
\begin{tabular}{@{\hskip7pt}c@{\hskip7pt}c@{\hskip7pt}c}
\hline
Sample             & $k_{T_{int}}$ & $k_{\theta_{j}}$ \\
&\\
\hline
Radio-bright GRBs  & $-1.3{^{-0.3}_{0.3}}$ & $-1.3{^{-0.2}_{0.3}}$ \\
& & \\
Radio-dark GRBs    & $-1.2{^{0.3}_{-0.6}}$  & $−1.6{^{-0.9}_{0.5}}$ \\ \hline
\end{tabular}
\end{table}

\begin{table}
\caption{List of GRBs showing VHE >100 MeV.}
\label{tab:vhe}
\begin{tabular}{ccccc}

\rowcolor[HTML]{FFFFFF} 
\hline
{\color[HTML]{000000} GRB}                             & {\color[HTML]{000000} $T_{90}$}   & {\color[HTML]{000000} $z$}                             & {\color[HTML]{000000} $E_{\gamma,iso}$}                           & {\color[HTML]{000000} Radio}                       \\
\hline \hline
\rowcolor[HTML]{F9F9F9} 
\cellcolor[HTML]{FFFFFF}{\color[HTML]{000000} 090323}  & {\color[HTML]{333333} 133}   & {\color[HTML]{333333} 3.57}                          & {\color[HTML]{333333} 4.10E+54}                         & \cellcolor[HTML]{FFFFFF}{\color[HTML]{000000} Yes} \\
\rowcolor[HTML]{FFFFFF} 
\cellcolor[HTML]{FFFFFF}{\color[HTML]{000000} 090328}  & {\color[HTML]{333333} 57}    & {\color[HTML]{333333} 0.736}                         & {\color[HTML]{333333} 1.00E+53}                         & \cellcolor[HTML]{FFFFFF}{\color[HTML]{000000} Yes} \\
\rowcolor[HTML]{F9F9F9} 
\cellcolor[HTML]{FFFFFF}{\color[HTML]{000000} 090902B} & {\color[HTML]{333333} ···}   & {\color[HTML]{333333} 1.883}                         & {\color[HTML]{333333} 3.09E+54}                         & \cellcolor[HTML]{FFFFFF}{\color[HTML]{000000} Yes} \\
\rowcolor[HTML]{FFFFFF} 
\cellcolor[HTML]{FFFFFF}{\color[HTML]{000000} 100414A} & {\color[HTML]{333333} 26}    & {\color[HTML]{333333} 1.368}                         & {\color[HTML]{333333} 7.79E+53}                         & \cellcolor[HTML]{FFFFFF}{\color[HTML]{000000} Yes} \\
\rowcolor[HTML]{FFFFFF} 
{\color[HTML]{000000} 131108A}                         & {\color[HTML]{000000} 19}    & {\color[HTML]{000000} 2.4}                           & {\color[HTML]{000000} 5.80E+53}                         & {\color[HTML]{000000} Yes}                         \\
\rowcolor[HTML]{FFFFFF} 
\cellcolor[HTML]{FFFFFF}{\color[HTML]{000000} 130907A} & {\color[HTML]{333333} 115}   & {\color[HTML]{333333} 1.238}                         & {\color[HTML]{333333} 3.30E+54}                         & \cellcolor[HTML]{FFFFFF}{\color[HTML]{000000} Yes} \\
\rowcolor[HTML]{FFFFFF} 
{\color[HTML]{000000} 160509A}                         & {\color[HTML]{000000} 371}   & {\color[HTML]{000000} 1.17}                          & {\color[HTML]{000000} 5.76E+53}                         & {\color[HTML]{000000} Yes}                         \\
\rowcolor[HTML]{FFFFFF} 
{\color[HTML]{000000} 151027A}                         & {\color[HTML]{000000} 124}   & {\color[HTML]{000000} 0.81}                          & {\color[HTML]{000000} 2.40E+52}                         & {\color[HTML]{000000} Yes}                         \\
\rowcolor[HTML]{FFFFFF} 
{\color[HTML]{000000} 180720B}                         & {\color[HTML]{000000} 48.9}  & \cellcolor[HTML]{FFFFFF}{\color[HTML]{2A2A2A} 0.654} & \cellcolor[HTML]{FFFFFF}{\color[HTML]{333333} 6.00E+53} & {\color[HTML]{000000} Yes}                         \\
\rowcolor[HTML]{FFFFFF} 
\cellcolor[HTML]{FFFFFF}{\color[HTML]{000000} 190829A} & {\color[HTML]{2A2A2A} 58.2}  & {\color[HTML]{2A2A2A} 0.0785}                        & {\color[HTML]{333333} 5.00E+50}                         & \cellcolor[HTML]{FFFFFF}{\color[HTML]{000000} Yes} \\
\rowcolor[HTML]{FFFFFF} 
\cellcolor[HTML]{FFFFFF}{\color[HTML]{000000} 190114C} & {\color[HTML]{2A2A2A} 361.5} & {\color[HTML]{2A2A2A} 0.4245}                        & {\color[HTML]{333333} 3.50E+53}                         & \cellcolor[HTML]{FFFFFF}{\color[HTML]{000000} Yes} \\
\hline
\vspace{1mm}
\end{tabular}

  
\end{table}










\bsp	
\label{lastpage}

\end{document}